\documentclass[lettersize,journal]{IEEEtran}
\usepackage{amsmath,amsfonts}
\usepackage{algorithmic}
\usepackage{algorithm}
\usepackage{array}
\usepackage[caption=false,font=normalsize,labelfont=sf,textfont=sf]{subfig}
\usepackage{textcomp}
\usepackage{stfloats}
\usepackage{url}

\usepackage{verbatim}
\usepackage{graphicx}
\usepackage{cite}
\usepackage{nomencl}
\usepackage{tabularx}
\makenomenclature

\begin{document}

\title{Flexible Transmission: A Comprehensive Review of Concepts, Technologies, and Market }

\author{Omid Mirzapour,~\IEEEmembership{Student Member,~IEEE},
        Xinyang Rui,~\IEEEmembership{Student Member,~IEEE},
        Brittany Pruneau,
        Mostafa Sahraei-Ardakani,~\IEEEmembership{Member,~IEEE}}
        % <-this % stops a space


\maketitle

\begin{abstract}
As global concerns regarding climate change are increasing worldwide, the transition towards clean energy sources has accelerated. Accounting for a large share of energy consumption, the electricity sector is experiencing a significant shift towards renewable energy sources. To accommodate this rapid shift, the transmission system requires major upgrades. Although enhancing grid capacity through transmission system expansion is always a solution, this solution is very costly and requires a protracted permitting process. The concept of flexible transmission encompasses a broad range of technologies and market tools that enable effective reconfiguration and manipulation of the power grid for leveraged dispatch of renewable energy resources. The proliferation of such technologies allows for enhanced transfer capability over the current transmission network, thus reducing the need for grid expansion projects. This paper comprehensively reviews flexible transmission technologies and their role in achieving a net-zero carbon emission grid vision. Flexible transmission definitions from different viewpoints are discussed, and mathematical measures to quantify grid flexibility are reviewed. An extensive range of technologies enhancing flexibility across the grid is introduced and explored in detail. The environmental impacts of flexible transmission, including renewable energy utilization and carbon emission reduction, are presented. Finally, market models required for creating proper incentives for the deployment of flexible transmission and regulatory barriers and challenges are discussed.

\end{abstract}


\nomenclature{$ATC$}{Available Transfer Capability}
% \nomenclature{$CBM$}{Capacity Benefit Margin}
\nomenclature{$TRM$}{Transmission Reliability Margin}
\nomenclature{$TTC$}{Total Transfer Capability}
\nomenclature{$\alpha$}{Power transfer increment factor}
\nomenclature{$\Omega_{FR}$}{Grid flexibility region}
\nomenclature{$\Omega_{AR}$}{Admissible region}
\nomenclature{$\Omega_{ESSR}$}{Extended steady-state security region}
\nomenclature{$\lambda_n$}{ Locational marginal price at bus $n$}
\nomenclature{$F_l$}{Active power flow on line \it{l}}
\nomenclature{$b_l$}{Susceptance of line \it{l}}
\nomenclature{$\theta_n$}{Voltage angle at bus \it{n}}
\nomenclature{$\varphi_l$}{Phase shifter angle installed on line \it{l}}
\nomenclature{$P_n$}{Active power injected into bus \it{n} }
\nomenclature{$Q_n$}{Reactive power injected into bus \it{n}}
\nomenclature{$V_n$}{Voltage magnitude at bus \it{n}}
\nomenclature{$N$}{Set of buses}
\nomenclature{$L$}{Set of lines}
\nomenclature{$NF$}{Network Flexibility index}
\nomenclature{$TF$}{Transmission Flexibility index}
\nomenclature{$BuSFI$}{Bulk System Flexibility Index}
\nomenclature{$S_l$}{Apparent power flow on line \it{l}}
\nomenclature{$PTDF$}{Power Transfer Distribution Factor}
\nomenclature{$L_{ik}$}{Load Shedding at bus \it{i} under outage scenario \it{k}}
% \nomenclature{$FTR$}{Financial Transmission Right}
% \nomenclature{$CR$}{Congestion Rent}
\nomenclature{$d_n$}{Demand at bus \it{n}}
\nomenclature{$s_l$}{Marginal value of susceptance on line \it{l}}
\nomenclature{$TC$}{Total generation cost}
\nomenclature{$\gamma_l$}{Partial switching on line \it{l}}
\printnomenclature

\section{Introduction}
\IEEEPARstart{T}{he} electricity sector has evolved rapidly over the past two decades with steep growth of production from renewable energy sources (RES). The reduced cost of harvesting renewable energy alongside the global push for a net-zero carbon electric grid has made renewable energy resources the prime alternative for electricity generation. Improvements in technology and economies of scale have resulted in levelized cost of energy (LCOE) from renewable energy resources to levels below that of even cheap coal-fired power plants \cite{IRENA,henze_2022}. Many countries, including the United States, have adopted targets for achieving a net-zero power grid ~\cite{us_plan,ger_plan,dk_plan,paris,energy.gov_2021}, which requires massive integration of renewable generation. The transition from fossil fuels to renewable generation faces many challenges, including intermittency, uncontrollability, and dependence on weather. Another major challenge is the transfer of renewable power, which requires substantial upgrades to the transmission system~\cite{navon2020integration,conlon2019assessing}.

%The cost of electricity from utility-scale solar farms has dropped by 85\% between 2010-2020~\cite{IRENA}. At the same time, as the impacts of climate change on the environment and human life are better sensed and understood, battling global warming has been gaining more force and momentum. Currently, 196 parties adopting the Paris Agreement are committed to achieving the climate-neutral grid goal by 2050~\cite{paris}. Replacing conventional fossil-fueled plants with renewable energy resources is one of the prime goals of this agreement. United States, Germany, and Denmark are among the countries that have set the goal of a carbon-free grid by 2050~\cite{us_plan,ger_plan,dk_plan}. United States Department of Energy recently released the blueprint for reaching a carbon-neutral power sector relying on 40\% penetration of solar energy~\cite{energy.gov_2021}. 

\begin{figure}[t]
\centering
\includegraphics[width=\columnwidth,keepaspectratio]{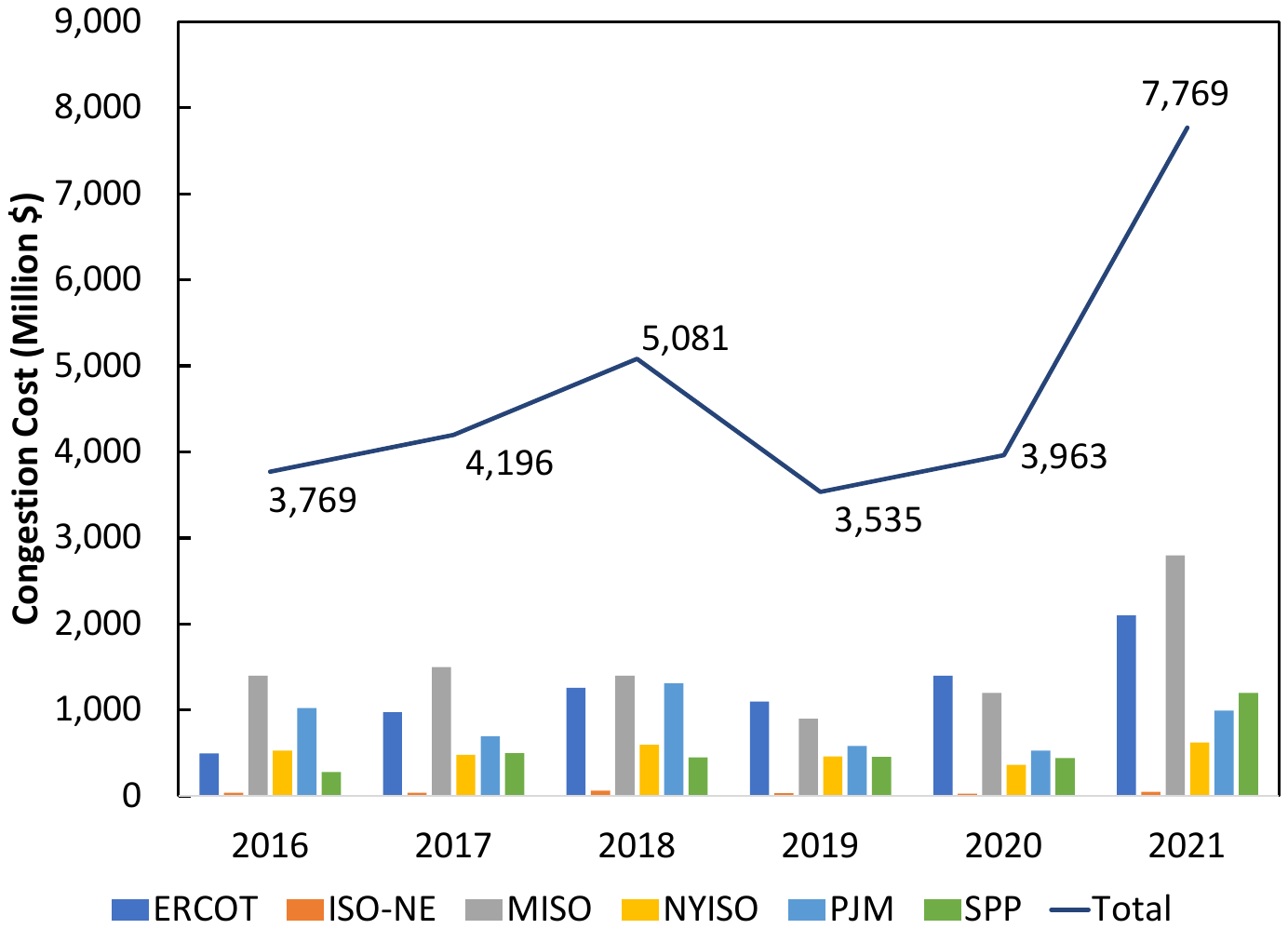}
\caption{Congestion cost in the U.S. by ISO/RTO 2016-2021}
\label{CRTO}
\end{figure}

%Integrating renewable energy resources into the power grid has introduced new challenges to the reliable operation of the system. Unlike conventional thermal plants, the system operator or plant owner does not schedule renewable energy resources output. Wind and solar generation outputs are volatile and demonstrate variations based on weather conditions. Additionally, increased penetration of inverter-based renewable energy resources, like solar power, has caused challenges for secure grid operation, including reduced system inertia and voltage deviations. 

%Ultimately, delivering uncertain renewable energy requires extra transfer capacity of the grid. 

Increased levels of congestion cost on the U.S. national grid are a clear indication of this necessity. Transmission congestion incurred over \$5 billion cost on the U.S. grid in 2020~\cite{doe2020national}. Fig.~\ref{CRTO} shows the reported annual congestion cost for six major ISO/RTOs during 2016-2021 except for CAISO, which does not publish any data on congestion cost~\cite{pjm, NYISO, miso,spp,ercot,isone}. These ISO/RTOs represent 58\% of the U.S. electricity consumption; therefore, the congestion cost is a good indicator of the increasing trend in congestion patterns across the U.S. The linear trend shows an average annual increase of \$507 million (17\%) in total congestion cost in the U.S. During 2020. The congestion cost dropped below the 2016 level due to the electricity consumption reduction influenced by the COVID-19 pandemic, which later increased with a sharp surge during 2021 to an unprecedented level of \$7.7 billion.

\begin{figure*}[t]
\centering
\includegraphics[width=\textwidth,keepaspectratio]{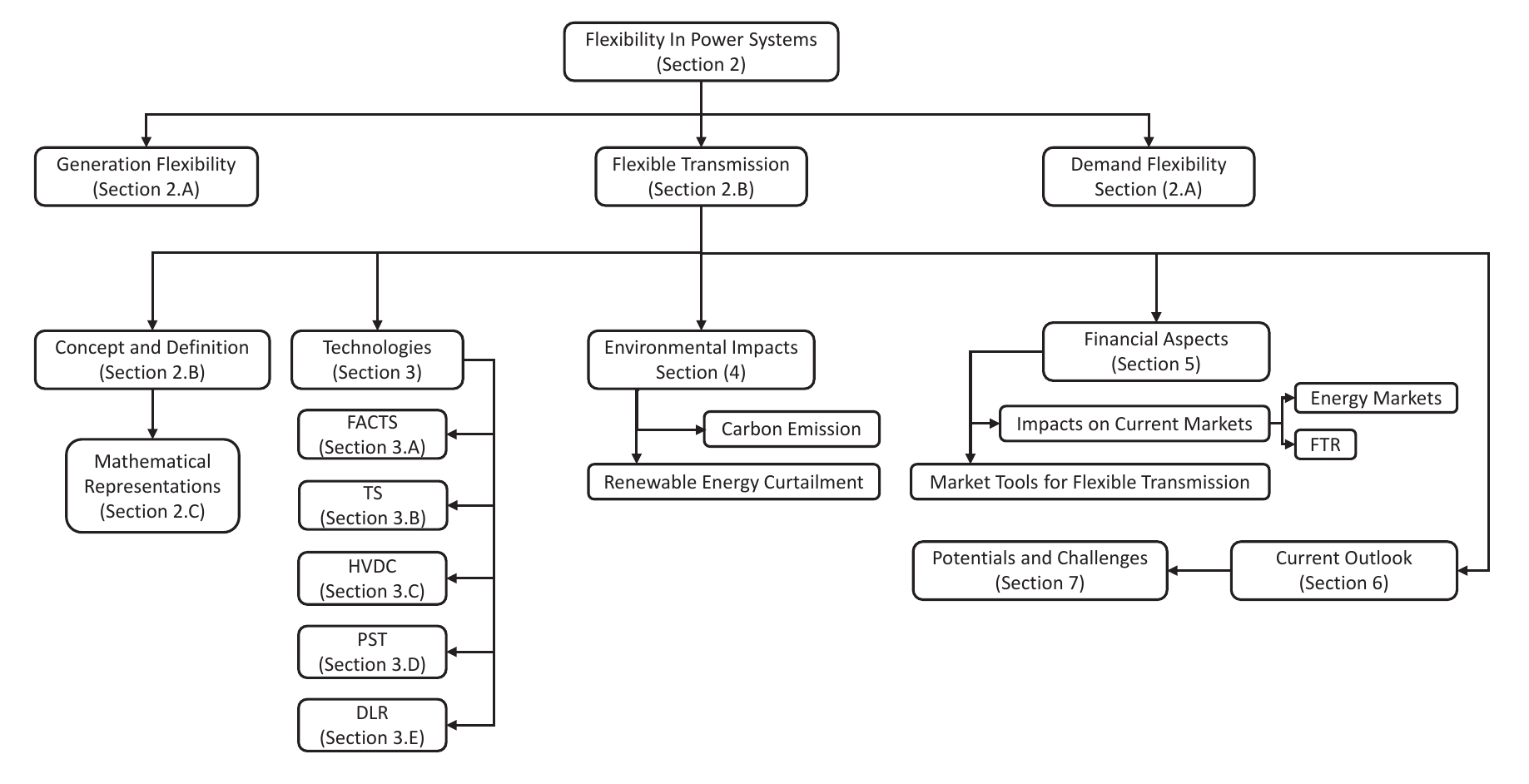}
\caption{Topics covered in this review paper}
\label{overview}
\end{figure*}

Studies by the National Renewable Energy Laboratory (NREL) show that a factor of 100-200\% should increase the capacity of the transmission network to accommodate the nation's renewable energy targets~\cite{cochran2015grid}. Energy System Integration Group (ESIG) transmission planning study suggests that the U.S. national grid requires doubling or tripling capacity to reach the carbon-free network vision~\cite{esig2020}. Transmission expansion projects, as the primary solution for this problem, require astronomical levels of investment above other problems, including siting and permitting barriers and lack of incentive for private investors. Alternatively, flexible transmission technologies can effectively enhance transfer capability over the grid and therefore enhance economic efficiency and reliability besides streamlining the integration of renewable energy resources. While building new transmission lines is still necessary, flexible transmission can offer a fast and cost-effective way of increasing transfer capability, which can reduce or postpone the need for new transmission.

Despite its high potential, there is a serious lack of comprehensive reviews in this area. To the best of our knowledge, the only review in this area is~\cite{li2018grid}, which investigates mathematical formulations and metrics for grid flexibility and computational methods to solve operation and planning problems in  the presence of flexible transmission. Some reviews have addressed flexible ac transmission system (FACTS) technologies, including~\cite{albatsh2015enhancing,eslami2012survey}. However, their scope is limited to FACTS technologies, while many options provide flexibility for the transmission network. This paper seeks to fill this gap by presenting an overview of available technologies enabling flexibility in the transmission system and discussing different aspects of flexible transmission, including financial and environmental factors. The main contribution of this paper is a detailed discussion of the following important items:
\begin{itemize}
    \item A consensual definition of flexible transmission based on the most recent developments,
    \item Mathematical tools for quantification and evaluation of transmission flexibility,
    \item Comprehensive review of technologies to obtain grid flexibility,
    \item Investigation of financial aspects of flexible transmission addressing the impact on current electricity markets and proper market tools for deployment,
    \item Environmental impacts of flexible transmission in renewable energy deployment and carbon emission,
    \item A comprehensive overview of current market and industry adoption of flexible transmission technologies,
    \item Identification of current barriers and suggestions for future research directions.
\end{itemize}
To this end, Section~\ref{definition} reviews the concept of flexibility in engineering and specifically in power systems. The section further addresses the challenges leading to transmission system flexibility and presents a definition of flexible transmission. Section~\ref{sources} introduces various technologies and methods enabling transmission system flexibility. Section~\ref{EI} discusses the impact of grid adjustments on carbon emission and renewable energy curtailment under different levels of renewable energy penetration and generation mix. Section~\ref{market} discusses current energy market structures and the impact of flexible transmission on both energy and financial transmission right (FTR) markets. Section~\ref{pt} addresses current barriers and prospects of flexible transmission. Finally, Section~\ref{conclude} summarizes the paper and presents final remarks and future research directions. The flow of this paper is conceptually visualized in Fig.~\ref{overview}.

\section{Transmission Flexibility Definition and Metrics}
\label{definition}
 This section presents an overview of flexibility in engineering systems, followed by the definition of power system flexibility and necessity in response to electricity system transformation. Accordingly, flexible transmission is introduced as a central part of power system flexibility, and the mathematical metrics for indicating transmission system flexibility are reviewed.

 \subsection{Flexibility in Power Systems}
 
 Flexibility is a broad concept used in various systems and applications from different points of view. Therefore, presenting a single and general-purpose definition of flexibility is not appropriate. Generally, flexibility is defined as the ability of a system to respond to a range of uncertain future states by taking an alternative course of action within an acceptable cost threshold and time window~\cite{zhao2015unified}. The International Energy Agency (IEA) defines flexibility as the general ability of a system to react to changes in generation and demand over time~\cite{challengeguide}. Considering the most general definitions presented above, flexibility can be studied from different aspects, including time horizon, type of variability, and sector. From the time horizon perspective, flexibility can be studied over short-term or long-term, ranging from milliseconds to months or years. Short-term analyses are performed for operational studies to ensure the system's reliable operation within security margins \cite{tovar2019generalized,poplavskaya2021making}. On the other hand, long-term flexibility analyses are performed in planning studies to ensure the system is robust against uncertainties in fuel prices or resource availability from an economic viewpoint \cite{ma2013evaluating}. Technical flexibility studies focus on the system's secure operation \cite{otashu2021cooperative}, while economic flexibility studies seek to employ flexible resources to maximize surplus \cite{brouwer2015operational}. Finally, the variability in flexibility studies can be predictable or involve uncertainty. Predicted variations include planned outage of transmission or generation units for maintenance, which can be implemented in the deterministic optimal power flow or unit commitment models \cite{xie2016reliability, phommixay2021two}. However, for uncertain variation, including renewable generation output variations and forced outages due to contingencies, stochastic or robust models may be required to ensure the secure operation of the grid \cite{yamujala2021stochastic, krommydas2022flexibility}. 
 
 Considering this framework, we can now review various definitions of power system flexibility in the literature. Reference ~\cite{lannoye2014transmission} defines power system flexibility as the ability of the system to deploy its resources to meet changes in net load. This definition takes a technical approach toward power system flexibility. The net load in this definition is the difference between the electricity demand and variable renewable generation, i.e., wind and solar. The steep growth in the share of RES has led to increased volatility in net load and subsequently increased demand for power system flexibility to accommodate higher levels of uncertainty. Another technical definition of power system flexibility focusing on generation flexibility is presented in \cite{dvorkin2014assessing}, which describes flexibility as the ability of power systems to provide enough ramping and capacity in response to demand variation on operational timescale. From the economic point of view, flexibility can be described as the system's competency to provide ample capacity in case of future uncertainties and variabilities at marginal cost \cite{bistline2018turn}. Risk management criteria determine the compromise between the cost of extra capacity and flexibility value in \cite{doe2020national}.
 
\begin{figure}[t]
\centering
\includegraphics[width=\columnwidth,keepaspectratio]{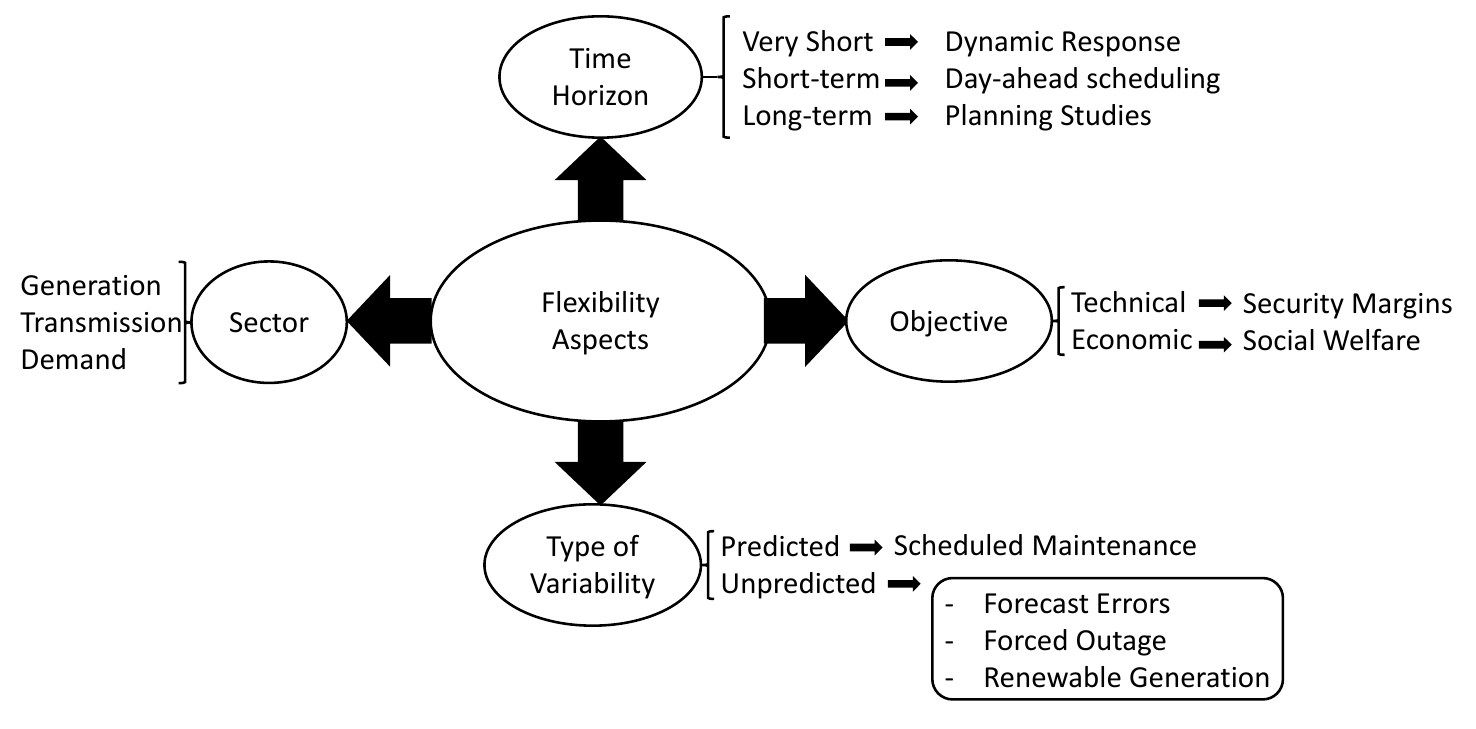}
\caption{Various aspect of flexibility in power system studies}
\label{aspects}
\end{figure}

 The most prevalent categorization of power system flexibility studies is by sector: generation, demand, and transmission flexibility. Generation flexibility is the most conventional area of power system flexibility studies. Fast response units are commonly dispatched to provide ramping requirements on the operational time horizons \cite{wu2014thermal,huertas2019hydropower}. The automatic generation control (AGC) is further utilized to compensate for mismatches between load and generation on shorter timescales \cite{li2015connecting,sajjadi2019governor}. Demand flexibility, on the other hand, seeks to exploit flexible loads to enhance the regulation capability of the power system to cope with uncertainties in the operation of the power system~\cite{CHEN2018125}. Demand Side Management (DSM) and Demand Response (DR) are the two well-known frameworks that make use of operational practices and guidelines, including shifting load to off-peak periods and consequently reducing generation cost and congestion in the network \cite{warren2014review, huang2019demand}. Multiple practices have been proposed in the literature that can be utilized to leverage flexibility on the demand side, including time-of-use tariffs, critical peak pricing, demand bidding and smart metering \cite{venizelou2018development, jabbari2017two, jang2015demand, chen2019energy, van2019smart}. Fig.~\ref{aspects} summarizes the various aspects of flexibility in power systems addressed in the literature. Reference ~\cite{degefa2021comprehensive} presents a comprehensive review of flexibility taxonomy and classification, considering the most recent publications.

 \subsection{Flexible Transmission Concept}
 
 Although there is a vast body of literature on generation and demand flexibility, transmission flexibility is more or less absent from flexibility studies. The reason is that the transmission network has historically provided sufficient capacity to deliver energy and accommodate standard reserve margins for reliability. Most of the existing literature in the power system flexibility area are centered around generation flexibility and  transmission system is considered a constraint in dispatching generation flexibility ~\cite{papalexopoulos2016impact}. However, the steep growth of renewable generation has pushed the legacy transmission network to its limits by introducing extra levels of uncertainty in the generation, new congestion patterns, and the need for security enhancements, including reactive power and voltage support requirements. Additionally, renewable generation provides lower flexibility levels than conventional thermal generation. Thus, higher penetration of renewable energy resources demands for enhanced operation of flexible transmission assets, i.e., FACTS devices, phase shifting transformers (PST), and circuit breakers, in providing flexibility at the grid level. 
 
 The current literature does not offer a consensual definition of transmission flexibility. A consensual definition with quantifiable metrics is crucial for stakeholders to evaluate their schedules and coordinate system-wide flexibility, ensuring effective information flow.  Reference~\cite{li2018grid} defines transmission system (grid-side) flexibility as the ability of a power network to deploy its flexible resources to cope with volatile changes in the power system state in operation. However, this definition is limited to operational flexibility and does not consider the long-term flexibility requirements of the transmission network. From the long-term perspective, \cite{Realoptionanalysis} defines flexibility in transmission investment as the ability of investors to choose among an extended range of alternatives, including postponement, abandonment, and operational flexibility enhancement under uncertain market environments. Enhancing operational flexibility helps the investor handle the long-term uncertainties and risks in the deregulated environment. It provides more secure decision-making, which reduces the investment risk for investors and provides more efficient assessment tools for regulatory entities.  

We define transmission system flexibility as the ability of the network to reconfigure itself in response to expected or unexpected variations over various time horizons to achieve maximum transfer capability from generation to demand. Accordingly, flexible transmission refers to various technologies and business practices that enable the transmission network to adjust to generation and demand uncertainties. Flexible transmission can alleviate congestion by enhancing transfer capability over the current grid and postponing grid expansion costs~\cite{amin2010securing}. A recent study by the Brattle Group over the Southwest Power Pool (SPP) grid shows that flexible transmission proliferation can leverage transfer capability by 100\%~\cite{brattle2021gridenhancing}. A recent case study by the U.S. Department of Energy on NYISO shows that grid-enhancing technologies, namely dynamic line rating and power flow controllers reduce renewable energy curtailment by 43\% and improve electricity rates for end-users, by alleviating congestion in the 2030 New York state grid vision with 70\% renewable energy penetration~\cite{usdoe-executive-summary}. Allocating such savings to different stakeholders in different market structures requires further research.

%%\begin{figure}[tb]
%%\centering
%%\includegraphics[width=\columnwidth,keepaspectratio]{Transmission Investment.pdf}
%%\caption{U.S. Grid Enhancement Investment 2014-2023}
%%\label{TI}
%%\end{figure}

%%\begin{figure}[tb]
%%\centering
%%\includegraphics[width=\columnwidth,keepaspectratio]{Energy_source.pdf}
%%\caption{Electricity Generation by Major Sources in the U.S. (2015-2021)}
%%\label{ES}
%%\end{figure}

%%\begin{figure}[tb]
%%\centering
%%\includegraphics[width=\columnwidth,keepaspectratio]{Renewable_Share.pdf}
%%\caption{Major Renewable Energy Generation in the U.S. (2015-2021)}
%%\label{RG}
%%\end{figure}

Considering the definition presented above, several aspects of flexible transmission must be addressed. Flexible transmission can be studied over various time horizons, from seconds to several years. Transmission networks can be considered in stability studies in response to contingencies~\cite{sahraei2015real}. Several studies have been conducted to deploy transmission network flexibility sources on operational time horizons, hourly to day-ahead operation~\cite{sang2017stochastic,khodaei2010transmissionuc}. Planning studies have incorporated flexible transmission to facilitate the integration of RER and prevent congestion in the long term~\cite{lu2005transmission,khodaei2010transmission}. We further need mathematical metrics to quantify the flexibility of the transmission network and the impacts of flexible transmission. Generally, grid flexibility can be evaluated from two points of view: technical and economic. From the technical point of view, various mathematical indices and geometric representations have been proposed in the literature. An overview of these metrics is provided in the following subsection. From the economic point of view, a lack of flexibility in transmission networks creates congestion in the grid, which results in an out-of-merit generation schedule and therefore incurs extra costs on consumers, which is formalized as congestion cost. A detailed analysis of congestion cost and how it can represent grid flexibility is provided in section \ref{market}.

\subsection{Technical Metrics for Transmission System Flexibility}

Available Transfer Capability (ATC) is one the most established metrics to evaluate the transmission network's ability to dispatch generated power to demand~\cite{hamoud2000assessment}. ATC measures the transmission system's capability to transfer extra power from the source to the sink. It is used for transmission expansion studies and operational constraints for day-ahead market scheduling \cite{bajrektarevic2006identifying}. In the U.S., FERC requires all ISO/RTOs to regularly calculate and report ATC to Open Access Same-time Information System (OASIS) for system upgrades and planning \cite{federal_energy_regulatory_commission_2020}. ATC is calculated based on Total Transfer Capability (TTC), Transmission Reliability Margin (TRM), and Capacity Benefit Margin (CBM) as follows \cite{mohammed2019available}:
\begin{equation}
    ATC = TTC - TRM - CBM.
\end{equation}

$TTC$ is the summation of maximum transfers allowable above base case power flow considering thermal, security, and contingency constraints. The calculation of TTC is the core of ATC calculation. This is a very important step, since overestimation of TTC puts the system at security risks and can lead to instability, while underestimation results in inefficient utilization of the transmission infrastructure and misleading signals for expansion \cite{karuppasamypandiyan2021day}. TTC calculation can be formulated using continuation power flow \cite{aman2014optimum}:
\begin{flalign}
&{\text{maximize }}  \alpha& \label{ttc}\\
&{\text{s.t.: }}  P_i = \sum_{j=1}^n V_iV_jY_{ij}\cos(\theta_j-\theta_i+\varphi_{ij}), && \forall I; \label{ttc4}\\
&\qquad Q_i = \sum_{j=1}^n V_iV_jY_{ij}\sin(\theta_j-\theta_i+\varphi_{ij}), && \forall i \in N;\label{ttc5}\\
&\qquad V_i^{\min} \leq V_i \leq V_i^{\max}, && \forall i \in N;\label{ttc6}\\
&\qquad \lvert S_{ij}\rvert \leq S_{ij}^{\max}, && \forall (i,j) \in L;\label{ttc7}\\
&\qquad P_i = P_i^0 (1+\alpha k_i), && \forall i \in N;\\
&\qquad Q_i = Q_i^0 (1+\alpha k_i), && \forall i \in N;\\
&\qquad TTC = \sum_{i\in\text{sink}} P_i(\alpha^{\max})-\sum_{i\in\text{sink}}P_i^0, \label{ttc9}
\end{flalign}
The mathematical problem seeks to maximize the additional increment in power transfer, $\alpha$, for each sink node subject to the network's physical constraints \eqref{ttc4}-\eqref{ttc7}. The total transfer capability is calculated as the total transfer amount above the base case power flow \eqref{ttc9}. $TRM$ is the transfer capability margin ISO reserves for coping with prevailing uncertainties, including renewable generation volatility, load forecast inaccuracies, and transmission contingencies. Maintaining sufficient TRM is necessary for reliable system operation and preventing curtailments \cite{nadia2020determination}. $CBM$ is allocated to ensure generation reserve deliverability in an interconnected system in case of a generation unit outage. The required amount of CBM is determined based on the historical value of loss of load expectation (LOLE) for each area. The areas with higher amounts of annual LOLE should maintain higher CBM margins for ensuring external reserve deliverability in case of generation outage \cite{mohammed2020capacity}.

To visualize grid-side flexibility, \cite{li2018grid} defines grid flexibility region $\Omega_{FR}$ as the difference between admissible region $\Omega_{AR}$ before and after implementing grid flexibility.
\begin{equation}
    \Omega_{FR}=\Omega_{AR}-\Omega_{AR0}.
\end{equation}

Admissible region $\Omega_{AR}$ used in the literature is defined as the area where all the flexibility resources can accommodate every uncertain power injection without violating feasibility constraints~\cite{chen2013steady,nguyen2018constructing}.
\begin{equation}
    \Omega_{AR} = \{(P_U,Q_U)|\forall (P_U,Q_U)\in \Omega_{AR}\\
    \exists (P_C,Q_C) \in \Omega_{ESSR}\},
\end{equation}
where $(P_U,Q_u)$ is the level of admissible uncertain active and reactive power injection that the grid can handle under the condition that ample flexibility from controllable power injection, e.g., generation control or FACTS devices, $(P_C,Q_C)$, is available, such that there exists a feasible solution under network physical constraints (3)-(6) described as extended steady-state security region ($\Omega_{ESSR}$).

To evaluate the flexibility of the network on the planning horizon, deterministic indices have been proposed in the literature. Reference \cite{bresesti2003power} evaluates the network flexibility under generation expansion scenarios based on the network capacity utilized by generation expansion.
\begin{equation}
    NF(S_{ij}^{max})=\frac{\sum_{ij \in N_l}S_{ij}^{max}(\sum_{k \in N_b}PTDF_k^{ij})}{\sum_{ij \in N_l}S_{ij}^{max}}.
\end{equation}

Network flexibility ($NF$) metric for each transmission capacity level ($S_{ij}^{max}$ is determined by the average usage of transmission capacity based on the total power transfer distribution factor (PTDF) for each injection, divided by total network capacity. The smaller value of NF means lower utilization of the network and, therefore, higher grid flexibility. 

A similar metric for transmission expansion planning has been proposed in \cite{zhao2009flexible}, based on the expected load shedding due to line outage for peak load under high  and low demand scenarios.
\begin{equation}
TF = E(\sum_{i\in N_b}\sum_{k\in N_l}L_{ik}P_k).    
\end{equation}

The transmission flexibility index ($TF$) is derived by calculating the expected value of load shedding at bus $i$ due to line $k$ outage for high and low peak demand scenarios. The smaller value of $TF$ shows smaller load shedding and, thus, higher flexibility of the transmission grid against line outage contingencies.

Reference \cite{capasso2014bulk} proposes bulk system flexibility index (BuSFI) as the maximum allowable variation in the generation at a node that minimizes power margin on the transmission system.
\begin{equation}
    \mathrm{minimize} \sum_{ij \in N_l} P_{ij}^{max}-|\sum_{k \in N_b}PTDF_k^{ij}(P_k^0+\Delta P_k)|
\end{equation}
s.t.: (3)-(6)
\begin{equation}
    BuSFI_k = \Delta P_k \qquad BuSFI_{tot} = \sum_{k \in N_gb} BuSFI_k.
\end{equation}

This formulation allows for calculating the local flexibility index, $BuSFI_k$ for each generation bus that allows generation expansion without violating transmission system security margins and total system flexibility index $BuSFI_{tot}$. This index offers a signal to generation owners about the ability of the transmission network to accommodate generation expansion at each area. The overview of various transmission system flexibility evaluation measures is summarized in Table~\ref{tab:flexmetric}.

\begin{table*}[tbh]
    \centering
    \caption{Transmission Flexibility Metrics}
    \label{tab:flexmetric}
    \renewcommand{\arraystretch}{1.2}
    \begin{tabularx}{\textwidth}{X X X X X}
        \hline\hline
         Flexibility Metric & Ref. No. &Time Horizon & Application & Computational Burden\\
         \hline
         Available Transfer Capability (ATC) & \cite{mohammed2019available} & Operation/Planning & Day-ahead Schedule/Maintenance Schedule/Expansion Studies  & Medium-High \\
         Admissible Region (AR) & \cite{li2018grid} & Operation & Visualization & High\\
         Network Flexibility (NF) & \cite{bresesti2003power} & Planning & Generation Expansion & Medium\\
         Transmission Flexibility (TF) & \cite{zhao2009flexible} & Planning & Transmission Expansion & Low-Medium\\
         Bulk System Flexibility Index (BuSFI) & \cite{capasso2014bulk} & Planning & Generation/Transmission Expansion & Medium-High\\
         \hline\hline
    \end{tabularx}
\end{table*}

\section{Sources of Transmission Flexibility}
\label{sources}

Transmission flexibility can be harnessed through various technologies and procedures. Prominent methods include  power flow control through discrete adjustment in grid topology~\cite{goldis2016shift} or continuous control of transmission line impedance~\cite{sahraei2015fast} or voltage phase~\cite{capitanescu2015enhanced}. This section presents an overview of each of the aforementioned technologies. 

\subsection{Flexible AC Transmission System (FACTS)}
\begin{figure}[t]
\centering
\includegraphics[width=\columnwidth,keepaspectratio]{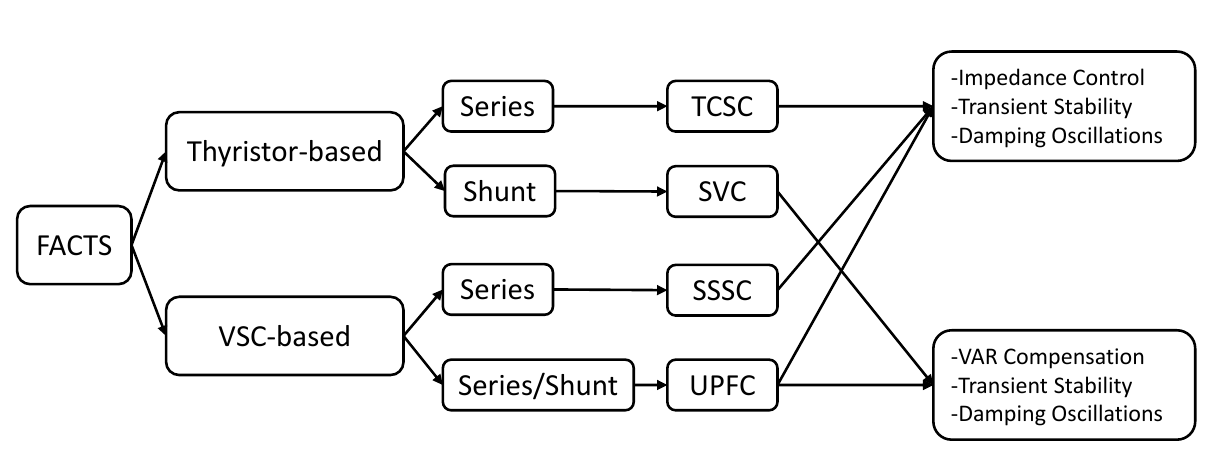}
\caption{FACTS Technology Overview}
\label{FACTS}
\end{figure}

FACTS technology intends to provide more efficient and reliable dispatch of electricity by providing tighter control over power flows in the transmission network~\cite{hingorani1993flexible}. FACTS devices can be deployed to enhance transient stability, voltage regulation, and mitigate system oscillations~\cite{gholipour2005improving,sayed2010all,zarghami2010nonlinear}. The first studies regarding the application of FACTS technology for enhancing transmission system flexibility and stability were carried out by Electric Power Research Institute (EPRI) in the U.S.~\cite{asare1994overview}. FACTS devices can control different properties of transmission elements, including line impedance, bus voltage magnitude, and voltage phase angle. FACTS devices can be categorized based on the controller technology. The older FACTS devices use thyristor switches to control line impedance or shunt capacitors. Thyristor-Controlled Series Compensator (TCSC) and Static VAR Compensators (SVC) are FACTS devices using thyristor controllers. In contrast, the more recent technology uses self-commutated Voltage Source Converters (VSC). Static Synchronous Series Compensators (SSSC) and Unified Power Flow Controllers (UPFC) are VSC-based FACTS devices. Based on their functionality, FACTS devices can be connected in series configuration (e.g., TCSC and SSSC) or shunt configuration (e.g., SVC), or a combination of both (e.g., UPFC). Generally, series compensators are used for line impedance and active power flow control~\cite{sahraei2015fast}, and shunt devices provide reactive power compensation. The UPFC is a more versatile device providing both shunt and series compensations~\cite{albatsh2015enhancing}. The recently developed distributed FACTS (D-FACTS) or modular FACTS (M-FACTS) provide the same functionality with the merit of modularity~\cite{kakkar2010recent,abdelsalam2014performance}. Figure~\ref{FACTS} summarizes FACTS classification based on controller, configuration, and functionalities.

In this paper, we focus on the power flow control functionalities of FACTS devices. Series FACTS devices can provide power flow control capabilities through line impedance adjustments, thus increasing the loading capability of the transmission system and alleviating congestion. FACTS technologies that can enhance transfer capability include TCSC, SSSC, and UPFC. Various types of FACTS devices use different techniques to alter the line reactance. TCSC devices can directly alter the line reactance, whereas UPFC and SSSC devices use a voltage injection to emulate reactance adjustments~\cite{rui2022linear, zhang2012flexible}. The first implementation of TCSC in Kayenta, Arizona, has shown a promising 30\% improvement in ATC~\cite{acharya2005facts}. Unlike TCSC, there has been few implementations of UPFC due to its high investment cost. The first practical UPFC implemented in Kentucky, U.S., consisted of two 160 MVA VSC-based units, which provided voltage and reactive power support and enhanced power flow capability across the region~\cite{paserba2004facts}.

The immediate economic benefit of series FACTS deployment in the transmission system is reducing the operational cost, as power flow control can enhance transfer capability and accommodate more generation from more cost-efficient units. Various existing literature has investigated the economic benefits of deploying FACTS devices and different levels of savings after solving power system operation problems with FACTS incorporated. In~\cite{sahraei2015fast} the two-stage linear optimal power flow with variable-impedance FACTS devices on IEEE 118-bus system shows up to 30\% cost saving compared to the base case when 40 lines are equipped with FACTS devices and 90\% of line reactance control is assumed.

Similarly, FACTS deployment can be an effective way of enhancing the integration of renewable generation. In~\cite{zhang2018optimal}, an optimal allocation problem for series FACTS devices and PSTs is solved. The results on the IEEE 118-bus system show that a 6.80\% increase in renewable generation can be accommodated with the deployment of TCSC and PST devices. Similarly, D-FACTS and FACTS deployment can lead to cost savings and help reduce renewable energy curtailment in most cases, according to the results in~\cite{sang2019effective}. It is worth noting that, as the results show in~\cite{sang2019effective},  the objective of power system operation models is to minimize the total operation cost; thus, FACTS deployment can lead to increased renewable curtailment in some cases. The effectiveness can be affected by the locations of FACTS devices and renewable generation, among other factors ~\cite{mirzapour2021environmental}.  

%% EDIT FROM HERE.
\subsection{Transmission Switching}
Transmission switching, also known as topology control, is an effective solution to enhance transfer capability. Transmission switching can be used for economic purposes or to enhance system reliability as a preventive or corrective action ~\cite{li2016real}. Northeast Power Coordinating Council has listed corrective transmission switching among Remedial Action Scheme (RAS) for avoiding abnormal voltage conditions~\cite{npcc}. Corrective transmission switching has been used for loss reduction, transmission line overloading relief, and improving power system reliability in the literature~\cite{hedman2011review,abdi2016corrective}. Although corrective transmission switching has been considered by ISO/RTOs for ad-hoc reconfiguration of the grid in response to contingencies, as in PJM's Special Protection Scheme (SPS)~\cite{pjmsps}, it is not yet integrated into energy or market management systems.

Transmission topology is traditionally assumed to be fixed, with the exception of planned or unplanned outages. Given that existing circuit breakers can open a line or transformer, network topology can be adjusted to deliver economic or reliability benefits under normal and contingency operation. Although the redundancy built into the transmission system ensures reliable power system operation in the long-term, it can lead to congestion in short-term operations. It is shown that reconfiguring the network by switching out specific lines under certain operational conditions will increase transfer capability. it is also well-known that switching out lightly loaded lines can alleviate voltage issues caused by such lines' capacitive components and enhance grid security. 

With the main goal of increasing the economic efficiency of power dispatch, optimal transmission switching has been used to enhance operational efficiency under normal operating conditions and decrease losses in the system through discrete adjustments in network topology~\cite{salkuti2018congestion}. The cost saving achieved by optimal transmission switching has shown up to 25\%~\cite{fisher2008optimal} in IEEE 118-bus system. However, the optimality of the solution is not guaranteed. Although transmission switching may seem to compromise network reliability by switching out transmission lines under normal operation, previous research has shown that co-optimizing transmission switching and generation dispatch saves cost while maintaining system reliability margins~\cite{henneaux2015probabilistic}. Co-optimizing network topology alongside N-1 optimal power flow shows that transmission switching can be implemented without endangering N-1 reliability criteria (system operation continuation under a single element outage)~\cite{khanabadi2012optimal}.
Despite the advantages that transmission switching introduces, its deployment has been rather limited. This is due to a number of reasons, including the computational burden of topology control problems, unpredictable distributional impacts on Locational Marginal Prices (LMP), and potential revenue inadequacy in FTR markets~\cite{hedman2011optimal}.

\subsection{Grid Interconnection and HVDC}

Traditionally, high-voltage AC (HVAC) systems have dominated the bulk power transmission grid, benefiting from the early development of high-voltage transformers that allow efficient power transmission over longer distances with minimal loss. However, with recent developments in DC technologies, including insulated-gate bipolar transistor (IGBT) valves into the power industry, high-voltage DC (HVDC) transmission increased its share of the transmission market. Although thyristor valves benefit from higher voltage and current ratings, IGBT valves due to self-commutation ability and controllability, are the primary option for VSC-HVDC applications~\cite{wang2016igbt}. After the first IGBT-based HVDC link project in 1997~\cite{shenai2015invention}, most VSC-HVDC projects relied on IGBT valves~\cite{danielsson2015transmission} and with expected 65\% share of VSC from HVDC market the IGBT is to become more prevalent~\cite{bnef2017electronhighways,usdoe2020advancedtransmission}. HVDC benefits from several technical and economic advantages, which makes it the superior alternative in many cases and even the sole option in particular applications. Currently, HVDC lines can deliver power over 3000 km while the maximum point-to-point distance for AC links is reported to be 1049 km~\cite{gu2018partial,shu2018research,chen2018variable}. This is due to the absence of reactive power components in HVDC links which minimizes the losses to resistive loss and allows full utilization of the line's capacity up to its thermal limits~\cite{long2007hvdc}. Further, HVDC offers complete power flow controllability, making it the prime option for delivering renewable generation over long distances, e.g., offshore wind generation~\cite{korompili2016review,sun2017renewable}. The controllability of power flow in HVDC link is crucial for smooth integration of intermittent renewable generation as it enables bidirectional power exchange between remote renewable resources and the AC grid while maintaining grid's stability and balance~\cite{watson2020overview}.
HVDC further provides grid flexibility by interconnecting regional transmission networks that are not synchronized. This will increase power supply security and enable interregional energy trading~\cite{wen2017enhancing}. The EU has set the target of 15\% capacity allocated for interconnection with neighboring markets by 2030~\cite{wendt20152030}. HVDC links are the only option for connecting asynchronous neighboring systems~\cite{rao2020frequency}. Finally, HVDC links can provide ancillary services, including power quality control and reactive power support~\cite{alassi2019hvdc}.

HVDC systems consist of two main blocks: interface to AC system and HVDC transmission. HVDC systems can be categorized based on the technologies and methods used in each block. AC system interface consists of ac-to-dc and dc-to-ac converters and transformers adjusting converter input/output level to AC transmission voltage level. The converter station, as the backbone of HVDC system, is designed based on either of two major technologies: line commutated converter (LCC) or voltage source converter (VSC). As the older technology, LCC stations operate based on the AC network configuration; therefore, they cannot participate in system cold start after blackouts~\cite{li2016research}. LCC further entails technical challenges, including excessive reactive power consumption and voltage reversal requirement at the DC terminal, which makes it the lesser attractive choice for the prospect of DC grids~\cite{li2015connecting}. The more recent technology, VSC is gradually dominating the HVDC links with its superior capabilities in providing ancillary service and power quality~\cite{xiong2020modeling}. VSC offers enhanced controllability over peripheral in hybrid grids and multi-terminal DC (MTDC) networks~\cite{rodriguez2017multi}.

Despite HVDC's advantages in power transmission, there are still challenges against deploying HVDC. The first challenge is the high cost of converter stations, especially the VSC technology ~\cite{xiang2016cost}. Second is the lack of cost-efficient DC circuit breakers, which has slowed the pace of DC grid deployment. DC cables' high cost and limited voltage ratings are other barriers to developing HVDC ~\cite{chen2017analysis}. Currently, the commercial rating for DC cables ranges from 200 to 1100 KV ~\cite{montanari2018criteria}. Regulatory hurdles and operational challenges further add to the complexity of operating HVDC lines, especially in the case of MTDC, where multiple ISO/RTOs are engaged. The issues regarding multiple ownership, MTDC control and management, and technical standardization of DC grids are the subject of future research. Table~\ref{tab:hvdc} provides a techno-economic comparison of HVDC versus HVAC transmission and HVDC technologies comparison.

\begin{table}[tb]
    \centering
    \caption{HVAC, LCC-HVDC, VSC-HVDC Comparison~\cite{rahman2021comparison}}
    \label{tab:hvdc}
    \renewcommand{\arraystretch}{1.2}
    \begin{tabularx}{\columnwidth}{X X X }
        \hline\hline
          & HVAC & HVDC\\
         \hline
         Conductor Cost & Higher ($\sim$\$1.5 Billion for 6000 MW, 765 kV, 2000 km) & Lower ($\sim$\$700 Million for 6000 MW, 800 kV, 2000 km) \\
         Substation Cost & Lower ($\sim$\$600 Million for 6000 MW, 765 kV, 2000 km) & Higher ($\sim$\$1 Billion for 6000 MW, 800 kV, 2000 km)\\
         Losses & Higher ($\sim$\$1 Billion for 6000 MW, 765 kV, 2000 km) & Lower ($\sim$\$500 Million for 6000 MW, 800 kV, 2000 km) \\
         Right of Way & Higher ($\sim$240 m for 6000 MW, 765 kV, 2000 km) & Lower ($\sim$90 m for 6000 MW, 800 kV, 2000 km) \\
         Maximum Span & Lower ($\sim$ 1049 km) & Higher ($\sim$ 3000 km)\\
         Protection Scheme & Mature and Cost-efficient & Developing and Expensive\\
         Power Flow Control & Lower & Higher\\
         Asynchronous Interconnection Capability & No & Yes\\
         \hline\hline
    \end{tabularx}
    \begin{tabularx}{\columnwidth}{X X X }
     & LCC-HVDC & VSC-HVDC\\
     \hline
     Cold-Start Capability & No & Yes\\
     Reactive Power Consumption & Higher & Lower\\
     Ancillary Service Procurement & Lower & Higher\\
     Converter Cost & Lower & Higher\\
     \hline\hline     
    \end{tabularx}
\end{table}

\subsection{Voltage Phase Control}
Voltage phase control over the transmission line allows active power flow rerouting throughout the AC transmission grid and can enhance the transfer capability. Two major technologies that allow voltage phase control over transmission lines include FACTS devices and phase-shifting transformers (PST). FACTS technology can provide voltage phase shift alongside other ancillary services. The most notable FACTS devices used for phase shift include thyristor-controlled phase shifters (TCPS), variable series capacitors (VSC), and unified power flow controllers (UPFC). FACTS phase shifters can provide dynamic response for transient stability and inter-regional oscillation mitigation~\cite{mihalivc2004transient,sajjadi2021svpwm}. While FACTS devices can offer more versatile control over transmission grid parameters, their investment cost is much higher than ordinary PST. PST can increase the transfer capability of the transmission system to provide power flow control by controlling the relative voltage phase angle difference between different areas of the grid and preventing parallel and loop flows relatively inexpensively~\cite{siemens_pst, huang2003feature}. PST can further be utilized to interconnect asynchronous systems and provide voltage regulation for enhancing supply security by adjusting the phase angle deviations at the point of interconnection that can be incurred by the frequency difference between the two systems~\cite{rasolomampionona2011interaction}. PST, however, lacks the dynamic response for providing stability control for the grid. Reference \cite{verboomen2008use} proposes an optimization framework for minimizing the risk of congestion by deploying PST on pivotal lines in the grid. This framework can further be utilized in broader planning studies where the requirements of transmission systems over longer time horizons need to be evaluated. Reference \cite{ippolito2004selection} proposes an index for prioritizing active power flow through lines using TCPS in a bilateral energy market. Reference \cite{urresti2020pre} proposes an optimization model for coordinated congestion management over interconnected RTOs using PST and introduces indices for cross-border PST site selection. Reference \cite{sidea2017sizing} proposes an optimal location framework for PST to maximize wind energy penetration in the network. The impact of pivotal PSTs (close to interconnection) on cross-border power transactions and flexibility provision in each TSO is further discussed in~\cite{korab2016impact}. PST has also been considered for enhancing the demand-side flexibility of plug-in electric vehicles (PEVs) in conjunction with tap-changing transformers~\cite{nikoobakht2019stochastic}. The 24-hour optimal power flow simulation on a modified IEEE 118-bus system with 5 wind farms, 4000 PEV fleet with stochastic availability and on-line tap-changing transformers and phase shifting transformers show that coordinating the PEV flexibility with transmission flexibility resources through linearized ac optimal power flow can achieve 11\% dispatch cost saving and 45\% reduction of wind spillage.

\subsection{Dynamic Line Rating}
Transmission line ratings were initially defined as reliability measures to ensure that power flow on the lines does not incur security risk over the grid~\cite{fernandez2016review}. Line ratings are often based on the thermal limits of conductors to prevent line overheating. However, these ratings are conventionally set rather conservatively by line owners, based on nameplate ratings with limited consideration of ambient temperature or wires condition. Nameplate rating is based on worst-case ambient assumptions and is not updated during the transmission line lifetime. Improving the methods for setting line ratings can increase efficiency in the transmission network and the entire power system on a larger scale~\cite{wallnerstrom2014impact}. With the growth of communication infrastructure and smart grid technology, it is now possible to set thermal limits using real-time weather data rather than the conservative ratings, the system operators use. This would allow more effective use of the current network capacity and enhance the transfer capacity of the grid~\cite{michiorri2015forecasting}. To consider more realistic assumptions regarding the ambient temperature, some utilities update line ratings for summer and winter temperatures. Some independent system operators, including PJM and ERCOT, use ambient adjusted ratings (AAR), which use online weather monitoring service data (including National Oceanic and Atmospheric Administration) to update line ratings based on atmospheric conditions~\cite{ferc2019dlr}.
This method calculates line ratings by step functions based on ambient data.

%%% EDIT From here. 
%%% 1) TALK ABOUT FERC REGULATION
%%% 2) TALK ABOUT HOW WIND IS THE MOST INFLUENTIAL FACTOR.

\begin{figure}[t]
\centering
\includegraphics[width=\columnwidth,keepaspectratio]{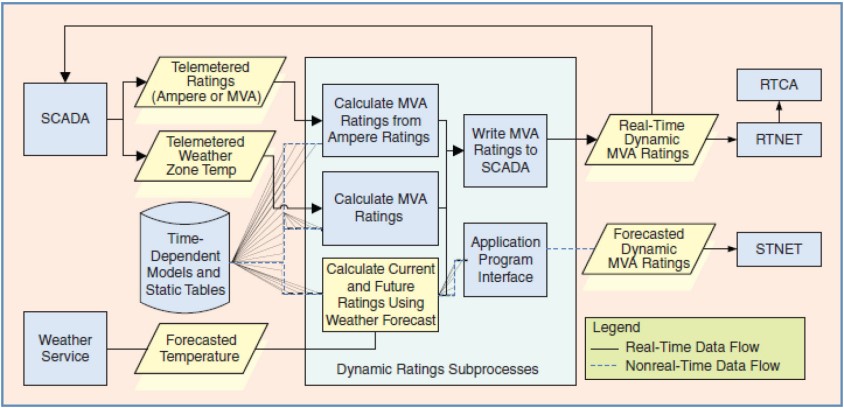}
\caption{DLR implementation in ERCOT's real-time EMS~\cite{hur2009high}}
\label{DLR}
\end{figure}

Dynamic Line Rating (DLR) considers the largest information set, including local weather conditions, wind speed and direction, solar irradiance, and line condition. Wind speed plays the most influential role among these factors as it can enhance line thermal capacity by 10\%-40\% through convective cooling~\cite{inl-dynamic-line-rating}. This potential can be best achieved by coupling line loading and cooling in areas with high penetration of wind generation~\cite{fernandez2016review}. Accordingly, DLR requires the most extensive equipment upgrades as well. This equipment includes sensors for measuring ambient conditions as well as line conditions. These sensors can be installed on the lines or use remote sensing (LIDAR). Line-based sensors provide more direct and accurate information compared to ground-based sensors. However, installing and maintaining these sensors require line outages, and they are more vulnerable to tampering. The next upgrade must be implemented on the communication infrastructure to transfer data to the energy management system (EMS). The communication medium can be cellular, satellite, microwave, or radio network. The choice of the communication medium depends on the location of sensors, the data's size, and the data transmission rate.  Fig.~\ref{DLR} shows one of the suggested DLR implementation schemes within ERCOT's EMS.

To effectively implement DLR, first, congested lines need to be identified based on the data provided by the system operator. Sensor placement depends on several factors, including line span and land diversity. Data communication must be implemented carefully since it is most susceptible to breaches and cybersecurity threats. The computational methods and coordination between DLR and more conventional line rating methods are important steps toward implementing DLR. Considering the intermittent nature of renewable energy resources and their interrelation with congestion in transmission systems, efficient forecasting methods need to be considered in rating calculation methods~\cite{michiorri2015forecasting}.  

Implementing DLR entails economic benefits, which include avoided transmission expansion investment cost on congested lines, reduced generation cost by alleviating congestion and out-of-merit dispatch of units, increased renewable generation utilization and reduced renewable energy spillage, and finally, increased system reliability, improved reserve deliverability and system disturbance management during summer. In~\cite{li2019day} DLR has been implemented in the day-ahead stochastic unit commitment model alongside transmission switching. The results on the RTS-79 system show that DLR can incur up to 20\% dispatch cost saving and 72\% wind energy curtailment reduction when co-optimized alongside network topology.
DLR enhances market competitiveness by enhancing market participants' access to the transmission system. DLR also improves transmission system reliability by increasing flexibility in line ratings compared to static line rating and providing more informed decision-making for operators and line owners rather than ad-hoc uprating.
A New York Power Authority (NYPA) study shows that dynamic line rating can increase line carrying capacity by 50\% compared to static rating~\cite{energy-dynamic-line-rating}. Another study by Oncor indicates that incorporating DLR into ERCOT's energy management system (EMS) increased wind utilization by 30-70\% compared to static rating and 6-14\% compared to AAR. Potomac Economics estimates a \$128 million saving in 2020 accrued by implementing AAR in MISO~\cite{potomac-economics-miso-report}. 

Currently, most system operators in the U.S., including CAISO, ISO-NE, MISO, NYISO, and SPP, use seasonal rating in compliance with FERC order 881, which is closest to static rating, and line ratings are rarely updated in contingency situations upon system operator request~\cite{caiso-managing-line-ratings,iso-ne-order-881,nyiso-ferc-order-881,spp-order-881-compliance}. Although DLR and AAR can effectively increase the operational efficiency of the grid, they are still not considered in the planning and long-term scheduling of the grid. Federal Energy Regulatory Commission (FERC) Order 1000 requires transmission planning models incorporating advanced technologies, including dynamic line rating~\cite{ferc-order-1000}. Table~\ref{tab:flextrans} reviews various sources of flexibility presented in this section.

\begin{table*}[tbh]
    \centering
    \caption{Flexible Transmission Technologies}
    \label{tab:flextrans}
    \renewcommand{\arraystretch}{1.2}
    \begin{tabularx}{\textwidth}{X X X X X}
        \hline\hline
         Flexibility Source & Control Parameter & Adjustment Mode & Implementation Cost & Operation Cost Saving \\
         \hline
         FACTS & Line Impedance/Voltage Phase and Magnitude & Continuous & Medium-High & up to 30\%~\cite{sahraei2015fast} \\
         TS & Grid Topology & Discrete & Low & up to 25\%~\cite{fisher2008optimal}\\
         HVDC & Capacity/Full Power Flow Control & Continuous & High & up to 50\% (loss reduction)~\cite{alassi2019hvdc}\\
         PST & Voltage Phase & Continuous & Medium & $\sim$12\%~\cite{nikoobakht2019stochastic}\\
         DLR & Capacity & Continuous & Low & $\sim$20\%~\cite{li2019day}\\
         \hline\hline
    \end{tabularx}
\end{table*}

\section{Environmental Impacts}
\label{EI}
Besides the economic benefits, flexible transmission can offer environmental advantages by contributing to the transition toward a carbon-free grid. Flexible transmission can accommodate intermittent renewable generation by adjusting the transfer capability accordingly and allow for higher levels of renewable energy penetration ~\cite{nikoobakht2019stochastic}. Accordingly, carbon emissions decline as fossil-fueled generation units are replaced by clean energy resources. Further considering transmission flexibility in the planning horizon leads to better generation and transmission expansion decisions. Various flexible transmission technologies have been studied for emission reduction. Reference \cite{sang2017stochastic} proposes variable-impedance FACTS devices co-optimization alongside a stochastic day-ahead unit commitment model to minimize wind energy curtailment. Transmission switching has been investigated in several studies to reduce carbon emissions. Reference \cite{shao2019low} proposes a minimum-emission optimal dispatch by adding emission cost to the dispatch cost function. This approach has been applied to transmission planning and joint generation and transmission expansion to minimize renewable energy spillage in the long term~\cite{sun2017analysis,seddighi2015integrated}. However, including emission costs in the objective function is not favored by current industry practices since it artificially increases total dispatch cost in the system and  leads to inefficient economic signals for investment. Multi-objective optimization has been used to address this shortcoming~\cite{lokeshgupta2018multi}. Although this approach introduces better economic signals, it is computationally burdensome for large networks, especially for transmission switching scheduling, which already introduces binary variables to the problem. Another challenge with multi-objective optimization is the challenges of picking a solution from the set of Pareto optimal solution. 

The literature has also addressed dynamic line rating to enhance renewable energy utilization~\cite{fernandez2016review}. The concerns regarding the power system reliability trade-off have been investigated in~\cite{park2018stochastic}, where the authors introduce an optimization framework to minimize wind spillage through dynamic line rating while maintaining reliability criteria within the acceptable limit. Joint optimization of multiple sources of transmission flexibility can help leverage the environmental benefits while covering the shortcomings of each source with the other one. Reference \cite{peker2018benefits} co-optimizes transmission switching and energy storage, leading to reduced wind curtailment and reduced load shedding. Dynamic line rating and transmission switching have also been utilized to minimize the dispatch cost while constraining carbon emissions within certain limits~\cite{jabarnejad2020facilitating}.

Although the literature suggests that flexible transmission leads to emission reduction and higher renewable energy utilization, under current market scheduling models with the primary objective of maximizing social welfare, flexible transmission  optimization may lead to adverse environmental impacts under certain operational conditions. This happens when the flexible transmission co-optimization model finds accommodating cheap thermal units such as coal more economical compared to congested renewable resources and chooses to curtail renewable energy and dispatch thermal generation. Such condition has been studied for variable-impedance FACTS devices~\cite{mirzapour2021environmental,mirzapour2023transmission}. Similar studies for transmission switching show similar results under current market structures~\cite{jabarnejad2020facilitating,peker2018benefits}.

\section{Financial Challenges and Market Implications}
\label{market}
The electricity sector was heavily regulated before the 1970s. However, deregulation starting with the generation side, showed that market-based operation brings higher efficiency to the electricity sector~\cite{ela2014evolution}. The distribution system has also evolved to some level of competition since then~\cite{wu2019pool}. The financial aspects of flexible transmission can be addressed from two points of view. The first aspect is the cost savings offered by flexible transmission through enhancing congestion management programs.
The second perspective is the impact each technology has on the stakeholders in the deregulated environment and the practical market tools for compensating flexible transmission owners based on their performance. Under the US's prevailing wholesale energy market structure, the generation units' payments are based on nodal LMPs. Transmission switching creates discrete adjustments to the transmission system topology, resulting in abrupt changes to LMPs. This impact conflicts with the competitive energy market and creates market power for specific stakeholders. The LMP alterations further disrupt FTR markets based on nodal price differences. The other issue regarding flexible transmission is that, unlike generation and demand, there is no competitive structure for flexible transmission owners which creates a lack of incentive for investors to participate in this sector. FTRs originally introduced to hedge transmission owner revenues against prevailing uncertainties. FTRs can be considered as congestion payments based on the LMP difference between the point of injection and the point of withdrawal as long as it passes the simultaneous feasibility test (SFT). SFT ensures that power flows resulting from these financial contracts do not violate network constraints. In this paradigm, each flexible transmission action including transmission switching, impedance control, or phase control can be modeled as a pair of injections. The owners receive FTRs based on the additional power flow they provide.~\cite{sahraei2018merchant} proposes a framework to incorporate merchant power flow controllers in a competitive market based on FTR allocation. FTRs are calculated as:
\begin{equation}
    FTR_l = F_l(\lambda_{l,to}-\lambda_{l,fr})
\end{equation}
where $\lambda$ is the locational marginal price at each end of the line and $F_l$ is the line flow obtained from the market economic dispatch solution. Total congestion revenue can be accordingly obtained:
\begin{equation}
    CR= \sum_n\lambda_n(d_n-\sum_{g\in G(n)}P_g^\ast)
\end{equation}
where $d_n$ is the demand at each bus and $P_g^\ast$ is the optimal output of generating unit at the bus. This paper suggests that compensating power flow controller (PFC) owners through FTR allocation proportional to additional congestion revenues they provide is adequate as long as the feasibility region is convex.

Another compensation method proposed in the literature is based on the marginal value of flexible transmission operation~\cite{sahraei2015transfer,sahraei2012active,sahraei2012market,sahraei2013marginal}. Calculating the flexible transmission's marginal value depends on the power flow controller type used. The marginal payments can be derived directly from the associated dual variables for flexible transmission technologies incurring continuous adjustments to power flow variables, including variable impedance FACTS devices and voltage phase controllers. Considering the power flow equation and its associated dual variable in DCOPF problem:
\begin{equation}
    F_l-b_l(\theta_{l,to}-\theta_{l,fr})= b_l(\varphi_l)\quad[s_l]
\end{equation}
the marginal value of susceptance adjustment can be calculated as:
\begin{equation}
    \frac{\partial TC}{\partial b_l}= s_l(\theta_{l,to}-\theta_{l,fr}-\varphi_l)
\end{equation}
and similarly, the marginal value of voltage phase control can be calculated as:
\begin{equation}
    \frac{\partial TC}{\partial \varphi_l}= b_ls_l.
\end{equation}
Calculating the marginal value for switching action is complicated due to the presence of integer variables in the power flow equation.~\cite{fuller2012fast} proposes a fictitious continuous variable to account for the partial switching of each line:
\begin{equation}
    F_l-b_l(1-\gamma_l)(\theta_{l,to}-\theta_{l,fr})= b_l(\varphi_l)\quad[s_l]
\end{equation}
Adding the constraint $\gamma_l=0$ the DCOPF model yields the same results as the original DCOPF. However, the dual variable associated with the latter constraint can be used as an estimation of the marginal value of switching a specific line. However, since switching is discrete, this estimation brings large inaccuracies in calculating the value of switching actions.

To set up an efficient market structure, first, the different payment structures should be compared by profitability, welfare improvement, and computational tractability.

%%% EDIT FROM HERE.

\section{Industry Adoption of flexible transmission technologies}
ISO/RTOs have adopted flexible transmission technologies, to a limited extent, to improve the efficiency, reliability, and security of the power system. Moreover, the benefits and potentials of flexible transmission technologies are acknowledged by multiple ISO/RTOs, and flexible transmission adoption is expected to grow in the near future. For example, PJM anticipates that deploying emerging technologies such as FACTS and DLR to utilize the existing grid is an important part of the solution to facilitate power system decarbonization~\cite{pjm_future_report}. However, the current implementation of flexible transmission is limited due to the challenges mentioned previously in this paper. Meanwhile, FERC has considered pushing for the adoption of flexible transmission technologies through regulation, with efforts in this direction including Order No. 1000~\cite{ferc-order-1000}, Order No. 881~\cite{ferc881web}, and the workshop discussing performance-based ratemaking approaches to facilitate the adoption of transmission technologies~\cite{fercworkshop}. Therefore, various discussions and proposals have been made to increase flexible transmission implementation by ISO/RTOs~\cite{pjm_problem_statement, isone_881_compliance}. Additionally, developments have been made by the industry in recent years for flexible transmission devices and solutions to allow more efficient deployments and operations. For example, the recently introduced distributed or modular FACTS devices (D-FACTS or M-FACTS) which are cheaper, smaller, and more flexible than conventional FACTS devices~\cite{rogers2008some, soroudi2021controllable}, as well as tools for optimizing DLR and transmission switching. In this section, we present a review of the current and planned implementation of flexible transmission technologies by ISO/RTOs, and related technology development in the industry. 
\subsection{FACTS Technologies}
Prominent shunt FACTS devices such as the SVC and the STATCOM are widely deployed in ISO/RTOs to provide a variety of functionalities~\cite{pjm_benefits, isone_regional_2021}. However, the adoption of series FACTS devices providing reactance adjustments, which is the focus of this paper, has been rather limited. Series reactors are deployed in ISO-NE, with some of them serving the purpose of power flow control for certain transmission lines~\cite{iso-ne_tranplan}. Series compensation and other technologies are considered  to alleviate thermal overloading in the area of CAISO with project proposals submitted by entities such as Smart Wires Inc.~\cite{caiso_tranplan}. Similarly, series capacitors are deployed by PJM to maintain reliability and provide operational flexibility during outages~\cite{pjm_teac}. However, there is no specific mention from the ISO/RTOs regarding whether the operation of these series compensation devices is optimized in operation models such as security-constrained unit commitment (SCUC). Additionally, it is unclear how and to what extent these resources are being utilized to alleviate congestion, reduce operation costs, and facilitate renewable generation integration. There have been deployments of series FACTS devices intended to solve transmission bottlenecks and help the integration of RES. An example of such deployments is the Marcy South series compensation project in NYISO, which was completed by the New York Power Authority (NYPA) and New York State Electric \& Gas (NYSEG) in 2016~\cite{nypa_mssc_press}. Series capacitors installed in this project can help enhance the transfer capability and facilitate delivery of power generated by wind and hydro units~\cite{doe_advanced_transmission, nypa_mssc_press}. However, whether the installed series capacitors are being dynamically controlled and adjusted in the operation models is unclear. 

General Electric (GE) currently offers series compensation systems that can increase the power transfer capability of existing or newly built transmission lines and have been installed in different locations worldwide, such as Texas and Vietnam, to increase transmission systems' reliability, stability, and power transfer capability~\cite{ge_seriescompensation}. Hitachi Energy produces TCSC devices that can enable rapid dynamic modulation of the inserted reactance~\cite{hitachi_tcsc}. Siemens also manufactures a variety of FACTS devices, including the fixed series capacitor and UPFC~\cite{siemens_facts}. The UPFC Plus devices by Siemens provide fast power flow control capabilities that can improve the utilization of existing transmission capacities~\cite{siemens_upfcplus}. Additionally, modular or distributed FACTS (M-FACTS or D-FACTS) have also emerged in the industry. The SmartValve manufactured by Smart Wires Inc. is a modular SSSC (M-SSSC) device that provides both inductive and capacitive transmission line reactance adjustments while providing more flexibility and scalability in deployments with its modular design~\cite{smartwires_smartvalve}. 

\subsection{Transmission switching}
ISO/RTOs have participated in multiple studies to explore the potential benefits of implementing transmission switching. A pilot project, in partnership with PJM, used historical PJM market data to study the cost savings achieved and implications on FTR by implementing transmission switching~\cite{Brattle_advancedtechnologies, ruiz_ferc_switching}. Case studies in SPP and ERCOT systems also have been carried out to demonstrate the effectiveness of transmission switching to reroute power flow to alleviate congestion and overloading~\cite{ruiz_ferc_switching_spp_ercot}. The studies also revealed that transmission switching can allow higher utilization of newly built high-capacity transmission lines. Regarding the actual implementation of transmission switching, PJM provides a list of potential transmission switching solutions that may be utilized for solving congestion issues~\cite{PJM_switching_solutions}. However, PJM states that the displayed transmission switching solutions are not guaranteed to be taken into action in market operations. 

The industry has developed software tools for implementing transmission switching. The topology optimization software by NewGrid can help system operators reroute power flows through transmission switching to reduce congestion~\cite{newgrid_arpae}. Results from the studies conducted between NewGrid and SPP show that topology  control can achieve \$18-44 million in market cost savings annually~\cite{ruiz_switching_spp_pilot}. NewGrid has conducted case studies with SPP,  NewGrid has identified ISO/RTOs, market participants, and transmission owners as their potential partners~\cite{newgrid_arpae}.

\subsection{Dynamic Line Rating}
Intending to improve power system efficiency, FERC Order No. 881 requires that organized market operators provide and maintain systems and procedures allowing transmission owners to implement DLR~\cite{ferc881web}. In response to the FERC order, ISO/RTOs have prepared compliances incorporating DLR in their transmission systems. Testing and implementation of DLR started before the issuing of the FERC order. NYPA demonstrated DLR technologies through the Smart Grid Demonstration Project and the results showed that up to 25\% additional capacity can be provided for system operations~\cite{doe_improving_dlr}. PJM conducted studies with American Electric Power (AEP) from 2016 to 2018 on two transmission lines to better understand the impact and potential benefits of DLR technology~\cite{pjm_benefits, pjm_dlr_webnews}. Pilot projects to test DLR have also been conducted by ERCOT and NYISO~\cite{ferc_dlr}. Additionally, ISO-NE achieved significant consumer savings to alleviate heavy congestion by increasing line ratings on transmission ties with New York in response to extreme weather conditions in 2018~\cite{doe_dlr}. FERC Order No. 881 specifies requirements and definitions that are different from historical ISO practices, which involve AAR periods and temperature settings~\cite{pjm_aar_setting}.

The industry has also developed numerous DLR technologies. For example, the LineRate Suite by LineVision is said to provide up to 40\% increase in transmission line capacity with accurate DLR calculations~\cite{linevision_dlr}. Additionally, Operato's software SUMO provides DLR algorithms and helps system operators optimize system operation~\cite{Sumo_operato, Sumo_smartwires}.
\subsection{PST}
PSTs are deployed by ISO/RTOs in various locations for power flow control. ISO-NE currently has PSTs at seven locations to control active (real) power flows on the transmission system within operating limits~\cite{iso-ne_tranplan}. An example of PST deployment by CAISO is the installation of two parallel PSTs at Imperial Valley~\cite{cook2018phase}. The operation of PSTs, along with other controllable transmission devices, are integrated into CAISO's security-constrained economic dispatch (SCED) and security-constrained unit commitment (SCUC), meaning that the optimal positions of the PSTs are determined by CAISO's market systems~\cite{caiso_business}. Similarly, the operation of PSTs in NYISO operation is optimized by SCUC, thus adjusting the PST schedules to help power delivery into congested areas~\cite{nyiso_par}. The Market Management System (MMS) of ERCOT is enhanced so that the tap settings of PSTs are automatically determined in the day-ahead market and the reliability unit commitment (RUC) to achieve efficient market solutions~\cite{hui2012wind}. 

PSTs are manufactured by companies such as GE, Hitachi Energy, and Siemens to provide power flow control solutions. The products can be used for applications including stability enhancement, overloading prevention, and increasing transmission capacity. 
\subsection{HVDC}
ISO/RTOs plan multiple transmission projects that involve HVDC. CAISO currently has two HVDC projects planned for reliability purposes and four others to facilitate access to wind generation, with examples being the TransWest Express Transmission Project and the SunZia transmission~\cite{caiso_tranplan_process}. Prominent HVDC projects for NYISO include the Clean Path New York (CPNY) and Champlain Hudson Power Express~\cite{nyiso_internal_controllable}. Both projects involve underground and underwater HVDC lines, thus providing benefits such as resilience enhancement and landscape protection~\cite{CPNY_HVDC, CHPE_HVDC}. NYISO is also developing market rules for internal controllable lines as the CPNY project is internal to the New York Control Area (NYCA)~\cite{nyiso_internal_controllable}. A prominent HVDC project involving PJM is the SOO Green Project carried out by Direct Connect Development Company, which can deliver low-cost renewable generation from MISO areas to customers in PJM~\cite{pjm_soogreen}.  PJM also established the High Voltage Direct Current Senior Task Force (HVDCSTF), which studied the potential of allowing HVDC converters to participate in PJM's capacity market~\cite{pjm_hvdcstf_report}. 

Prominent companies in the industry have made advances in HVDC technologies. GE manufactures both LLC and VSC, which have been deployed across the world~\cite{ge_hvdc_brochure}. Hitachi Energy also provides VSC, LLC, and HVDC control systems that can be utilized for various applications~\cite{HVDC_hitachi}. 
\subsection{Summary}
Overall, flexible transmission technologies have experienced different deployment patterns. They have been adopted or planned to enhance transmission capacity and facilitate access to renewable generation. However, incorporating the operation of flexible transmission devices in the existing power system operation models and changing the market structure to allow merchant flexible transmission market participation is essential to fully harness the benefits of flexible transmission technologies. It appears that state-of-the-art operation models can only optimize PST setpoints, although developments to model other flexible transmission technologies are currently underway. 

\section{Potentials and Challenges}
\label{pt}
Implementing flexible transmission technologies is currently limited due to technical challenges and regulatory barriers. This section summarizes these challenges.

\subsection{Regulatory Barriers and Lack of Incentive}
The transmission system is considered a natural monopoly and, thus, is tightly regulated. Under such presumptions, the joint generation and expansion planning by a non-profit regulatory body, ensuring that transmission expansion projects procure technical requirements of the grid and is aligned towards welfare improvement for all stakeholders, has shown to be the superior solution by the literature~\cite{alrasheedi2023unit}. Flexible transmission, however, does not pose the characteristics of bulk transmission. Since the investment is smaller for these assets  than transmission lines, a competitive environment can be created by incorporating small investors in the ancillary service market or auction revenue rights market~\cite{opgrand2019role}. The regulatory barriers discourage flexible transmission deployment, although the technology is already available and mature. Currently, the regulated rate of return payment scheme does not provide sufficient incentive for small investors to deploy flexible transmission since the owner's economic benefits of flexible transmission operation are not fully received~\cite{kuosmanen2020capital}. Efficient market design can provide the opportunity for the proliferation of flexible transmission and reduce the need for costly and time-consuming transmission expansion projects. Market mechanisms further provide the environment for active participation of flexible transmission in power system operation and result in the efficient operation of transmission network~\cite{aazami2012comprehensive}. This will ultimately lead to the transition to a more reliable grid and streamline renewable energy resource integration.

The proposed models should follow the market theories to deregulate the transmission grid similar to generation and demand effectively. For this purpose, economically-efficient price signals should be introduced to ensure market efficiency through satisfying revenue adequacy for investment, social welfare improvement, and computational tractability. Three market designs are proposed to incorporate flexible transmission into current electricity markets. First is the Financial Transmission Rights (FTR) market. This market mechanism only works for market desings with nodal market designs as it requires spacial distribution of  locational marginal prices to compensate FTR owners based on congestion rent. The congestion rent cannot be calculated in single market price mechanisms as the spacial price data is unavailable. For such cases as in European market design a market-based compensation has been proposed using the cost-based redispatch signal prices~\cite{staudt2021merchant}. The strength of this method compared to the FTR based markets is that the investor profit is aligned with total welfare maximization. This means that any investment that increases the social welfare of the system will receive positive compensation from market clearing. This means that the cost-based redispatch compensation mechanism intrinsically overcomes the revenue adequacy problem in FTR markets~\cite{joskow2020competition,schulte2020vision}. The third compensation method is the marginal value compensation of each transmission project. The main obstacle against this method is that it provides no hedging for the investor against the probable risks in system operation and might lead to negative revenue in some cases. Some research have also introduced mixed regulated and merchant transmission investment that can be fit into current market mechanisms~\cite{matschoss2019german}. 

\subsection{EMS Integration and Computational Challenges}
Implementing flexible transmission technologies in the EMS and coordination with current procedures and emerging technologies, including energy storage systems, in a computationally efficient manner, is another aspect of future grid vision~\cite{marques2022grid}. Power system equations are nonlinear and non-convex, making short-term and long-term optimization models computationally demanding, especially for large-scale systems~\cite{haaberg2019fundamentals,lumbreras2016new}. Co-optimizing the power flow controller's placement and operation further adds to this complexity~\cite{sang2017stochastic}. Considering these complexities and the limited computational power available, current markets use simplified models with linearized equations for power flow. To ensure the reliable operation of the electricity market, the impact of intrinsic nonlinearities and power flow controllers on market efficiency should be identified. In case of a large mismatch, out-of-market corrections should be introduced.

 A prominent example of the increased computational burden caused by optimizing flexible transmission operation is the nonlinearity of the DC power flow equation resulting from considering the operation of variable-impedance FACTS devices such as TCSC.  Previous studies have proposed algorithms and models to efficiently co-optimize FACTS operation with power system operation and planning~\cite{rui2022linear,sahraei2015fast,sahraei2016computationally}. Optimal transmission switching adds to the computational burden by introducing integer variables that yield a mixed-integer problem with higher computational demand. Efficient methods for solving AC and DC formulated co-optimization of transmission switching and optimal power flow are available in the literature~\cite{kocuk2017new,bai2016two}. Efficient formulations and solution methods that are capable of addressing the nonlinearities of flexible transmission and acceptable solution time are interesting subjects of future studies.

\subsection{Technical Challenges}
Although flexible transmission technologies have developed significantly during the past two decades, some technical issues still hinder their deployment. In HVDC interconnections, the DC circuit breaker technology is not mature and cost-effective compared to its AC counterpart. The LCC converter stations, the more dominant converter technology, lacks the capability for weak AC grid support. Some of the studies in the research have proposed hybrid LCC/VSC-based links that benefit both VSC controllability and LCC cost-effectiveness. On the other hand, the VSC technology suitable for multi-peripheral DC networks is not cost-effective. Active circuit breakers are gaining attention for grid reconfiguration and HVDC link fault currents. Flexible converters and novel configurations are necessary to provide flexibility in AC and DC networks.

\subsection{Communication Infrastructure and Smart Grid}
Effective scheduling and optimization of flexible transmission assets require enhanced data flow and information exchange systems for proper coordination. Smart grid infrastructure is essential for effectively controlling flexible transmission, which requires upgrades in the power system's current communication schemes. The role of the smart grid is also significant in long-term decision-making and flexibility planning. The recent FLEXITRANSTORE pilot project has been developed to enhance smart grid requirements for leveraging efficient flexibility to achieve low carbon power systems across Europe~\cite{vita2021evaluating}. The importance of Information and Communication Technologies (ICT) becomes more significant as the decentralization trends increase in electricity markets increase and cross-border transactions are required to ensure supply security. Managing a multitude of uncertainties and sources of flexibility requires state-of-the-art communication schemes. The role of the smart grid in managing demand-side flexibility in a competitive market environment has been shown in~\cite{eid2015aggregation}. A similar scheme is missing in flexible transmission literature and should be addressed in future research. The final aspect of the smart grid's importance in flexibility is flexibility coordination between different sectors, especially demand-side and transmission flexibility schedules. As more distributed generation units are deployed in the distribution system, and demand response programs are increasing, scheduling these sources of flexibility with the transmission system is crucial and needs efficient information exchange between ISO/TSOs and Distribution System Operators (DSO)~\cite{schachter2016critical}.

% add references here.  

\section{Conclusion}
\label{conclude}
This paper offers a comprehensive review of flexible transmission, including the concept, mathematical representations, technologies, environmental issues, and market prospects. The concept of flexibility in transmission networks was conceptually explained and further illustrated with mathematical indices and visualization. The technologies that can be utilized to enhance the flexibility of the grid are listed and deeply reviewed, addressing both merits and shortcomings of each technology. The role of flexible transmission in the electricity sector decarbonization is addressed by each technology and operational condition that can lead to the adverse impact of flexible asset adjustments on carbon emission and renewable energy curtailment described. The integration of flexible transmission into current electricity markets is studied, addressing both impacts of flexible transmission on energy and FTR markets and market tools for incentivizing flexible transmission investment. Finally, current barriers against flexible transmission deployment and future research roadmap are envisioned. The main conclusions drawn from this study are listed below:
\begin{itemize}
    \item flexible transmission can offer considerable cost savings and reliability enhancement
    \item Although flexible transmission leverages the environmental benefits of renewables in most cases, under certain market conditions, it can lead to increased levels of carbon emission
    \item there are still some technical challenges in various areas of flexible transmission that need to be addressed, including DC circuit breakers and VSC technology
    \item under the current market environment, the flexible transmission investment is not compensated on a competitive basis, and the distributional impacts of flexible transmission on LMP and FTR markets need to be addressed
    \item Development of computational tools that can balance the computational time and optimal solution for the nonlinear flexible transmission co-optimization is required
\end{itemize}
  
\bibliographystyle{IEEEtran}
\bibliography{ref.bib}

\begin{thebibliography}{100}
\providecommand{\url}[1]{#1}
\csname url@samestyle\endcsname
\providecommand{\newblock}{\relax}
\providecommand{\bibinfo}[2]{#2}
\providecommand{\BIBentrySTDinterwordspacing}{\spaceskip=0pt\relax}
\providecommand{\BIBentryALTinterwordstretchfactor}{4}
\providecommand{\BIBentryALTinterwordspacing}{\spaceskip=\fontdimen2\font plus
\BIBentryALTinterwordstretchfactor\fontdimen3\font minus \fontdimen4\font\relax}
\providecommand{\BIBforeignlanguage}[2]{{%
\expandafter\ifx\csname l@#1\endcsname\relax
\typeout{** WARNING: IEEEtran.bst: No hyphenation pattern has been}%
\typeout{** loaded for the language `#1'. Using the pattern for}%
\typeout{** the default language instead.}%
\else
\language=\csname l@#1\endcsname
\fi
#2}}
\providecommand{\BIBdecl}{\relax}
\BIBdecl

\bibitem{IRENA}
IRENA, \emph{Renewable Power Generation Costs in 2020}.\hskip 1em plus 0.5em minus 0.4em\relax International Renewable Energy Agency, Abu Dhabi, 2021.

\bibitem{henze_2022}
\BIBentryALTinterwordspacing
V.~Henze, ``Cost of new renewables temporarily rises as inflation starts to bite,'' Jun 2022. [Online]. Available: \url{https://about.bnef.com/blog/cost-of-new-renewables-temporarily-rises-as-inflation-starts-to-bite}
\BIBentrySTDinterwordspacing

\bibitem{us_plan}
{United States Department of State}, ``The long-term strategy of united states: Pathways to net-zero greenhouse gas emissions by 2050,'' Nov 2021.

\bibitem{ger_plan}
{German Advisory Council on Environment}, ``Pathways towards a 100 \% renewable electricity system chapter 10: Executive summary and recommendations,'' Jan 2011.

\bibitem{dk_plan}
{Regulation of the European Parliament and of the Council on the Governance of the Energy Union and Climate Action}, ``Denmark’s integrated national energy and climate plan,'' Dec 2019.

\bibitem{paris}
UNFCCC, ``Paris agreement,'' \emph{United Nations Treaty Collection, Chapter XXVII 7. d}, 2018.

\bibitem{energy.gov_2021}
\BIBentryALTinterwordspacing
{United States Department of Energy}, ``Solar futures study,'' Sep 2021. [Online]. Available: \url{https://www.energy.gov/eere/solar/solar-futures-study}
\BIBentrySTDinterwordspacing

\bibitem{navon2020integration}
A.~Navon, P.~Kulbekov, S.~Dolev, G.~Yehuda, and Y.~Levron, ``Integration of distributed renewable energy sources in israel: Transmission congestion challenges and policy recommendations,'' \emph{Energy Policy}, vol. 140, p. 111412, 2020.

\bibitem{conlon2019assessing}
T.~Conlon, M.~Waite, and V.~Modi, ``Assessing new transmission and energy storage in achieving increasing renewable generation targets in a regional grid,'' \emph{Applied Energy}, vol. 250, pp. 1085--1098, 2019.

\bibitem{doe2020national}
{United States Department of Energy}, ``National electric transmission congestion study,'' 2020.

\bibitem{pjm}
\BIBentryALTinterwordspacing
{Monitoring Analytics}, ``State of the market report for {PJM},'' 2016-2021. [Online]. Available: \url{https://www.monitoringanalytics.com/reports/PJM_State_of_the_Market}
\BIBentrySTDinterwordspacing

\bibitem{NYISO}
\BIBentryALTinterwordspacing
{Potomac Economics}, ``State of the market report for the {NYISO} electricity markets,'' 2016-2021. [Online]. Available: \url{https://www.potomaceconomics.com/wp-content/uploads}
\BIBentrySTDinterwordspacing

\bibitem{miso}
\BIBentryALTinterwordspacing
------, ``State of the market report for the {MISO} electricity markets,'' 2016-2021. [Online]. Available: \url{https://www.potomaceconomics.com/wp-content/uploads}
\BIBentrySTDinterwordspacing

\bibitem{spp}
\BIBentryALTinterwordspacing
{SPP Market Monitoring Unit}, ``State of the market,'' 2016-2021. [Online]. Available: \url{https://www.spp.org/documents}
\BIBentrySTDinterwordspacing

\bibitem{ercot}
\BIBentryALTinterwordspacing
{Potomac Economics}, ``State of the market report for the {ERCOT} electricity markets,'' 2016-2021. [Online]. Available: \url{https://www.potomaceconomics.com/wp-content/uploads}
\BIBentrySTDinterwordspacing

\bibitem{isone}
\BIBentryALTinterwordspacing
{ISO-NE Internal Market Monitor}, ``Annual market report,'' 2016-2021. [Online]. Available: \url{https://www.iso-ne.com/static-assets/documents}
\BIBentrySTDinterwordspacing

\bibitem{cochran2015grid}
J.~Cochran, P.~Denholm, B.~Speer, and M.~Miller, ``Grid integration and the carrying capacity of the us grid to incorporate variable renewable energy,'' National Renewable Energy Lab.(NREL), Golden, CO (United States), Tech. Rep., 2015.

\bibitem{esig2020}
A.~Bloom, L.~Azar, J.~Caspary, N.~Miller, A.~Silverstein, J.~Simonelli, and R.~Zavadil, ``Transmission planningfor 100\% clean electricity,'' \emph{Energy System Integration Group}, 2021.

\bibitem{li2018grid}
J.~Li, F.~Liu, Z.~Li, C.~Shao, and X.~Liu, ``Grid-side flexibility of power systems in integrating large-scale renewable generations: A critical review on concepts, formulations and solution approaches,'' \emph{Renewable and Sustainable Energy Reviews}, vol.~93, pp. 272--284, 2018.

\bibitem{albatsh2015enhancing}
F.~M. Albatsh, S.~Mekhilef, S.~Ahmad, H.~Mokhlis, and M.~Hassan, ``Enhancing power transfer capability through flexible ac transmission system devices: a review,'' \emph{Frontiers of Information Technology \& Electronic Engineering}, vol.~16, no.~8, pp. 658--678, 2015.

\bibitem{eslami2012survey}
M.~Eslami, H.~Shareef, A.~Mohamed, and M.~Khajehzadeh, ``A survey on flexible ac transmission systems (facts),'' \emph{Organ}, vol.~1, p.~12, 2012.

\bibitem{zhao2015unified}
J.~Zhao, T.~Zheng, and E.~Litvinov, ``A unified framework for defining and measuring flexibility in power system,'' \emph{IEEE Transactions on power systems}, vol.~31, no.~1, pp. 339--347, 2015.

\bibitem{challengeguide}
\BIBentryALTinterwordspacing
{International Energy Agency}, ``Harnessing variable renewables: A guide to the balancing challenge,'' May 2011. [Online]. Available: \url{https://www.oecd.org/publications/harnessing-variable-renewables-9789264111394-en.htm}
\BIBentrySTDinterwordspacing

\bibitem{tovar2019generalized}
C.~Tovar-Ram{\'\i}rez, C.~Fuerte-Esquivel, A.~M. Mares, and J.~S{\'a}nchez-Gardu{\~n}o, ``A generalized short-term unit commitment approach for analyzing electric power and natural gas integrated systems,'' \emph{Electric Power Systems Research}, vol. 172, pp. 63--76, 2019.

\bibitem{poplavskaya2021making}
K.~Poplavskaya, J.~Lago, S.~Str{\"o}mer, and L.~De~Vries, ``Making the most of short-term flexibility in the balancing market: Opportunities and challenges of voluntary bids in the new balancing market design,'' \emph{Energy Policy}, vol. 158, p. 112522, 2021.

\bibitem{ma2013evaluating}
J.~Ma, V.~Silva, R.~Belhomme, D.~S. Kirschen, and L.~F. Ochoa, ``Evaluating and planning flexibility in sustainable power systems,'' in \emph{2013 IEEE power \& energy society general meeting}.\hskip 1em plus 0.5em minus 0.4em\relax IEEE, 2013, pp. 1--11.

\bibitem{otashu2021cooperative}
J.~I. Otashu, K.~Seo, and M.~Baldea, ``Cooperative optimal power flow with flexible chemical process loads,'' \emph{AIChE Journal}, vol.~67, no.~4, p. e17159, 2021.

\bibitem{brouwer2015operational}
A.~S. Brouwer, M.~van~den Broek, A.~Seebregts, and A.~Faaij, ``Operational flexibility and economics of power plants in future low-carbon power systems,'' \emph{Applied Energy}, vol. 156, pp. 107--128, 2015.

\bibitem{xie2016reliability}
K.~Xie, H.~Zhang, and C.~Singh, ``Reliability forecasting models for electrical distribution systems considering component failures and planned outages,'' \emph{International journal of electrical power \& energy systems}, vol.~79, pp. 228--234, 2016.

\bibitem{phommixay2021two}
S.~Phommixay, M.~L. Doumbia, and Q.~Cui, ``A two-stage two-layer optimization approach for economic operation of a microgrid under a planned outage,'' \emph{Sustainable Cities and Society}, vol.~66, p. 102675, 2021.

\bibitem{yamujala2021stochastic}
S.~Yamujala, P.~Kushwaha, A.~Jain, R.~Bhakar, J.~Wu, and J.~Mathur, ``A stochastic multi-interval scheduling framework to quantify operational flexibility in low carbon power systems,'' \emph{Applied Energy}, vol. 304, p. 117763, 2021.

\bibitem{krommydas2022flexibility}
K.~F. Krommydas, C.~N. Dikaiakos, G.~P. Papaioannou, and A.~C. Stratigakos, ``Flexibility study of the greek power system using a stochastic programming approach for estimating reserve requirements,'' \emph{Electric Power Systems Research}, vol. 213, p. 108620, 2022.

\bibitem{lannoye2014transmission}
E.~Lannoye, D.~Flynn, and M.~O'Malley, ``Transmission, variable generation, and power system flexibility,'' \emph{IEEE Transactions on Power Systems}, vol.~30, no.~1, pp. 57--66, 2014.

\bibitem{dvorkin2014assessing}
Y.~Dvorkin, D.~S. Kirschen, and M.~A. Ortega-Vazquez, ``Assessing flexibility requirements in power systems,'' \emph{IET Generation, Transmission \& Distribution}, vol.~8, no.~11, pp. 1820--1830, 2014.

\bibitem{bistline2018turn}
J.~E. Bistline, ``Turn down for what? the economic value of operational flexibility in electricity markets,'' \emph{IEEE Transactions on Power Systems}, vol.~34, no.~1, pp. 527--534, 2018.

\bibitem{wu2014thermal}
H.~Wu, M.~Shahidehpour, A.~Alabdulwahab, and A.~Abusorrah, ``Thermal generation flexibility with ramping costs and hourly demand response in stochastic security-constrained scheduling of variable energy sources,'' \emph{IEEE Transactions on Power Systems}, vol.~30, no.~6, pp. 2955--2964, 2014.

\bibitem{huertas2019hydropower}
D.~Huertas-Hernando, H.~Farahmand, H.~Holttinen, J.~Kiviluoma, E.~Rinne, L.~S{\"o}der, M.~Milligan, E.~Ibanez, S.~M. Martinez, E.~G{\'o}mez-L{\'a}zaro \emph{et~al.}, ``Hydropower flexibility for power systems with variable renewable energy sources: An iea task 25 collaboration,'' \emph{Advances in Energy Systems: The Large-scale Renewable Energy Integration Challenge}, pp. 385--405, 2019.

\bibitem{li2015connecting}
N.~Li, C.~Zhao, and L.~Chen, ``Connecting automatic generation control and economic dispatch from an optimization view,'' \emph{IEEE Transactions on Control of Network Systems}, vol.~3, no.~3, pp. 254--264, 2015.

\bibitem{sajjadi2019governor}
M.~Sajjadi and H.~Seifi, ``Governor parameter estimation considering upper/lower production limits,'' in \emph{2019 IEEE Milan PowerTech}.\hskip 1em plus 0.5em minus 0.4em\relax IEEE, 2019, pp. 1--6.

\bibitem{CHEN2018125}
Y.~Chen, P.~Xu, J.~Gu, F.~Schmidt, and W.~Li, ``Measures to improve energy demand flexibility in buildings for demand response (dr): A review,'' \emph{Energy and Buildings}, vol. 177, pp. 125--139, 2018.

\bibitem{warren2014review}
P.~Warren, ``A review of demand-side management policy in the uk,'' \emph{Renewable and Sustainable Energy Reviews}, vol.~29, pp. 941--951, 2014.

\bibitem{huang2019demand}
W.~Huang, N.~Zhang, C.~Kang, M.~Li, and M.~Huo, ``From demand response to integrated demand response: Review and prospect of research and application,'' \emph{Protection and Control of Modern Power Systems}, vol.~4, no.~1, pp. 1--13, 2019.

\bibitem{venizelou2018development}
V.~Venizelou, N.~Philippou, M.~Hadjipanayi, G.~Makrides, V.~Efthymiou, and G.~E. Georghiou, ``Development of a novel time-of-use tariff algorithm for residential prosumer price-based demand side management,'' \emph{Energy}, vol. 142, pp. 633--646, 2018.

\bibitem{jabbari2017two}
M.~Jabbari~Zideh and S.~S. Mohtavipour, ``Two-sided tacit collusion: Another step towards the role of demand-side,'' \emph{Energies}, vol.~10, no.~12, p. 2045, 2017.

\bibitem{jang2015demand}
D.~Jang, J.~Eom, M.~G. Kim, and J.~J. Rho, ``Demand responses of korean commercial and industrial businesses to critical peak pricing of electricity,'' \emph{Journal of Cleaner Production}, vol.~90, pp. 275--290, 2015.

\bibitem{chen2019energy}
Y.~Chen, S.~Mei, F.~Zhou, S.~H. Low, W.~Wei, and F.~Liu, ``An energy sharing game with generalized demand bidding: Model and properties,'' \emph{IEEE Transactions on Smart Grid}, vol.~11, no.~3, pp. 2055--2066, 2019.

\bibitem{van2019smart}
P.~Van~Aubel and E.~Poll, ``Smart metering in the netherlands: What, how, and why,'' \emph{International Journal of Electrical Power \& Energy Systems}, vol. 109, pp. 719--725, 2019.

\bibitem{degefa2021comprehensive}
M.~Z. Degefa, I.~B. Sperstad, and H.~S{\ae}le, ``Comprehensive classifications and characterizations of power system flexibility resources,'' \emph{Electric Power Systems Research}, vol. 194, p. 107022, 2021.

\bibitem{papalexopoulos2016impact}
A.~Papalexopoulos, C.~Hansen, R.~Frowd, A.~Tuohy, and E.~Lannoye, ``Impact of the transmission grid on the operational system flexibility,'' in \emph{2016 Power Systems Computation Conference (PSCC)}.\hskip 1em plus 0.5em minus 0.4em\relax IEEE, 2016, pp. 1--10.

\bibitem{Realoptionanalysis}
R.~Pringles, F.~Olsina, and F.~Garcés, ``Power transmission investment under uncertainty: A real option framework,'' in \emph{2015 18th International Conference on Intelligent System Application to Power Systems (ISAP)}, 2015, pp. 1--7.

\bibitem{amin2010securing}
S.~M. Amin, ``Securing the electricity grid,'' \emph{The Bridge}, vol.~40, no.~1, pp. 19--20, 2010.

\bibitem{brattle2021gridenhancing}
\BIBentryALTinterwordspacing
{Brattle Group}, ``Unlocking the queue with grid enhancing technologies,'' 2021. [Online]. Available: \url{https://www.brattle.com/wp-content/uploads/2021/06/21200_unlocking_the_queue_with_grid_enhancing_technologies.pdf}
\BIBentrySTDinterwordspacing

\bibitem{usdoe-executive-summary}
\BIBentryALTinterwordspacing
{U.S. Department of Energy}, ``{Grid Enhancing Technologies - A Case Study on Ratepayer Impact - February 2022},'' {U.S. Department of Energy}, 2022, {Executive Summary}. [Online]. Available: \url{{https://www.energy.gov/sites/default/files/2022-04}}
\BIBentrySTDinterwordspacing

\bibitem{sahraei2015real}
M.~Sahraei-Ardakani, X.~Li, P.~Balasubramanian, K.~Hedman, and M.~Abdi-Khorsand, ``Real-time contingency analysis with transmission switching on real power system data,'' \emph{IEEE Transactions on Power Systems}, vol.~31, no.~3, pp. 2501--2502, 2015.

\bibitem{sang2017stochastic}
Y.~Sang, M.~Sahraei-Ardakani, and M.~Parvania, ``Stochastic transmission impedance control for enhanced wind energy integration,'' \emph{IEEE Transactions on Sustainable Energy}, vol.~9, no.~3, pp. 1108--1117, 2017.

\bibitem{khodaei2010transmissionuc}
A.~Khodaei and M.~Shahidehpour, ``Transmission switching in security-constrained unit commitment,'' \emph{IEEE Transactions on Power Systems}, vol.~25, no.~4, pp. 1937--1945, 2010.

\bibitem{lu2005transmission}
M.~Lu, Z.~Dong, and T.~Saha, ``Transmission expansion planning flexibility,'' in \emph{2005 International Power Engineering Conference}.\hskip 1em plus 0.5em minus 0.4em\relax IEEE, 2005, pp. 893--898.

\bibitem{khodaei2010transmission}
A.~Khodaei, M.~Shahidehpour, and S.~Kamalinia, ``Transmission switching in expansion planning,'' \emph{IEEE Transactions on Power Systems}, vol.~25, no.~3, pp. 1722--1733, 2010.

\bibitem{hamoud2000assessment}
G.~Hamoud, ``Assessment of available transfer capability of transmission systems,'' \emph{IEEE Transactions on Power systems}, vol.~15, no.~1, pp. 27--32, 2000.

\bibitem{bajrektarevic2006identifying}
E.~Bajrektarevic, S.~Kang, V.~Kotecha, S.~Kolluri, M.~Nagle, S.~Datta, M.~Papic, J.~Useldinger, P.~Patro, L.~Hopkins \emph{et~al.}, ``Identifying optimal remedial actions for mitigating violations and increasing available transfer capability in planning and operations environments,'' in \emph{CIGRE, Paris, France}, 2006.

\bibitem{federal_energy_regulatory_commission_2020}
\BIBentryALTinterwordspacing
F.~E. R.~C. (FERC), ``Order no. 889,'' Aug 2020. [Online]. Available: \url{https://www.ferc.gov/industries-data/electric/industry-activities/open-access-transmission-tariff-oatt-reform/history-of-oatt-reform/order-no-889-1}
\BIBentrySTDinterwordspacing

\bibitem{mohammed2019available}
O.~O. Mohammed, M.~W. Mustafa, D.~S.~S. Mohammed, and A.~O. Otuoze, ``Available transfer capability calculation methods: A comprehensive review,'' \emph{International Transactions on Electrical Energy Systems}, vol.~29, no.~6, p. e2846, 2019.

\bibitem{karuppasamypandiyan2021day}
M.~Karuppasamypandiyan, P.~A. Jeyanthy, D.~Devaraj, and V.~A.~I. Selvi, ``Day ahead dynamic available transfer capability evaluation incorporating probabilistic transmission capacity margins in presence of wind generators,'' \emph{International Transactions on Electrical Energy Systems}, vol.~31, no.~1, p. e12693, 2021.

\bibitem{aman2014optimum}
M.~Aman, G.~Jasmon, A.~Bakar, and H.~Mokhlis, ``Optimum network reconfiguration based on maximization of system loadability using continuation power flow theorem,'' \emph{International journal of electrical power \& energy systems}, vol.~54, pp. 123--133, 2014.

\bibitem{nadia2020determination}
A.~Nadia, A.~H. Chowdhury, E.~Mahfuj, M.~S. Hossain, K.~Z. Islam, and M.~I. Rahman, ``Determination of transmission reliability margin using ac load flow,'' \emph{AIMS Energy}, vol.~8, no.~4, pp. 701--720, 2020.

\bibitem{mohammed2020capacity}
O.~O. Mohammed, M.~W. Mustafa, M.~N. Aman, S.~Salisu, and A.~O. Otuoze, ``Capacity benefit margin assessment in the presence of renewable energy,'' \emph{International Transactions on Electrical Energy Systems}, vol.~30, no.~9, p. etep12502, 2020.

\bibitem{chen2013steady}
S.~J. Chen, Q.~X. Chen, Q.~Xia, and C.~Q. Kang, ``Steady-state security assessment method based on distance to security region boundaries,'' \emph{IET Generation, Transmission \& Distribution}, vol.~7, no.~3, pp. 288--297, 2013.

\bibitem{nguyen2018constructing}
H.~D. Nguyen, K.~Dvijotham, and K.~Turitsyn, ``Constructing convex inner approximations of steady-state security regions,'' \emph{IEEE Transactions on Power Systems}, vol.~34, no.~1, pp. 257--267, 2018.

\bibitem{bresesti2003power}
P.~Bresesti, A.~Capasso, M.~Falvo, and S.~Lauria, ``Power system planning under uncertainty conditions. criteria for transmission network flexibility evaluation,'' in \emph{2003 IEEE Bologna Power Tech Conference Proceedings,}, vol.~2.\hskip 1em plus 0.5em minus 0.4em\relax IEEE, 2003, pp. 6--pp.

\bibitem{zhao2009flexible}
J.~H. Zhao, Z.~Y. Dong, P.~Lindsay, and K.~P. Wong, ``Flexible transmission expansion planning with uncertainties in an electricity market,'' \emph{IEEE Transactions on Power Systems}, vol.~24, no.~1, pp. 479--488, 2009.

\bibitem{capasso2014bulk}
A.~Capasso, A.~Cervone, M.~Falvo, R.~Lamedica, G.~Giannuzzi, and R.~Zaottini, ``Bulk indices for transmission grids flexibility assessment in electricity market: A real application,'' \emph{International Journal of Electrical Power \& Energy Systems}, vol.~56, pp. 332--339, 2014.

\bibitem{goldis2016shift}
E.~A. Goldis, P.~A. Ruiz, M.~C. Caramanis, X.~Li, C.~R. Philbrick, and A.~M. Rudkevich, ``Shift factor-based scopf topology control mip formulations with substation configurations,'' \emph{IEEE Transactions on Power Systems}, vol.~32, no.~2, pp. 1179--1190, 2016.

\bibitem{sahraei2015fast}
M.~Sahraei-Ardakani and K.~W. Hedman, ``A fast lp approach for enhanced utilization of variable impedance based facts devices,'' \emph{IEEE Transactions on Power Systems}, vol.~31, no.~3, pp. 2204--2213, 2015.

\bibitem{capitanescu2015enhanced}
F.~Capitanescu, ``Enhanced risk-based scopf formulation balancing operation cost and expected voluntary load shedding,'' \emph{Electric Power Systems Research}, vol. 128, pp. 151--155, 2015.

\bibitem{hingorani1993flexible}
N.~G. Hingorani, ``Flexible ac transmission,'' \emph{IEEE spectrum}, vol.~30, no.~4, pp. 40--45, 1993.

\bibitem{gholipour2005improving}
E.~Gholipour and S.~Saadate, ``Improving of transient stability of power systems using upfc,'' \emph{IEEE Transactions on power delivery}, vol.~20, no.~2, pp. 1677--1682, 2005.

\bibitem{sayed2010all}
M.~A. Sayed and T.~Takeshita, ``All nodes voltage regulation and line loss minimization in loop distribution systems using upfc,'' \emph{IEEE Transactions on power electronics}, vol.~26, no.~6, pp. 1694--1703, 2010.

\bibitem{zarghami2010nonlinear}
M.~Zarghami, M.~L. Crow, and S.~Jagannathan, ``Nonlinear control of facts controllers for damping interarea oscillations in power systems,'' \emph{IEEE Transactions on Power Delivery}, vol.~25, no.~4, pp. 3113--3121, 2010.

\bibitem{asare1994overview}
\BIBentryALTinterwordspacing
P.~Asare, T.~Diez, A.~Galli, E.~O'Neill-Carillo, J.~Robertson, and R.~Zhao, ``An overview of flexible ac transmission systems,'' \emph{ECE Technical Reports}, 1994. [Online]. Available: \url{http://docs.lib.purdue.edu/ecetr/205}
\BIBentrySTDinterwordspacing

\bibitem{kakkar2010recent}
V.~Kakkar and N.~Agarwal, ``Recent trends on {FACTS} and {D-FACTS},'' in \emph{2010 Modern Electric Power Systems}.\hskip 1em plus 0.5em minus 0.4em\relax IEEE, 2010, pp. 1--8.

\bibitem{abdelsalam2014performance}
A.~A. Abdelsalam, H.~A. Gabbar, and A.~M. Sharaf, ``Performance enhancement of hybrid {AC/DC} microgrid based {D-FACTS},'' \emph{International Journal of Electrical Power \& Energy Systems}, vol.~63, pp. 382--393, 2014.

\bibitem{rui2022linear}
X.~Rui, M.~Sahraei-Ardakani, and T.~R. Nudell, ``Linear modelling of series facts devices in power system operation models,'' \emph{IET Generation, Transmission \& Distribution}, vol.~16, no.~6, pp. 1047--1063, 2022.

\bibitem{zhang2012flexible}
X.-P. Zhang, C.~Rehtanz, and B.~Pal, \emph{Flexible AC transmission systems: modelling and control}.\hskip 1em plus 0.5em minus 0.4em\relax Springer Science \& Business Media, 2012.

\bibitem{acharya2005facts}
N.~Acharya, A.~Sode-Yome, and N.~Mithulananthan, ``Facts about flexible ac transmission systems (facts) controllers: practical installations and benefits,'' in \emph{Australasian universities power engineering conference (AUPEC), Australia}.\hskip 1em plus 0.5em minus 0.4em\relax Citeseer, 2005, pp. 533--538.

\bibitem{paserba2004facts}
J.~J. Paserba, ``How facts controllers benefit ac transmission systems,'' in \emph{IEEE Power Engineering Society General Meeting, 2004.}\hskip 1em plus 0.5em minus 0.4em\relax IEEE, 2004, pp. 1257--1262.

\bibitem{zhang2018optimal}
X.~Zhang, D.~Shi, Z.~Wang, B.~Zeng, X.~Wang, K.~Tomsovic, and Y.~Jin, ``Optimal allocation of series facts devices under high penetration of wind power within a market environment,'' \emph{IEEE Transactions on power systems}, vol.~33, no.~6, pp. 6206--6217, 2018.

\bibitem{sang2019effective}
Y.~Sang and M.~Sahraei-Ardakani, ``Effective power flow control via distributed facts considering future uncertainties,'' \emph{Electric Power Systems Research}, vol. 168, pp. 127--136, 2019.

\bibitem{mirzapour2021environmental}
O.~Mirzapour and M.~Sahraei-Ardakani, ``Environmental impacts of power flow control with variable-impedance facts,'' in \emph{2020 52nd North American Power Symposium (NAPS)}.\hskip 1em plus 0.5em minus 0.4em\relax IEEE, 2021, pp. 1--6.

\bibitem{li2016real}
X.~Li, P.~Balasubramanian, M.~Sahraei-Ardakani, M.~Abdi-Khorsand, K.~W. Hedman, and R.~Podmore, ``Real-time contingency analysis with corrective transmission switching,'' \emph{IEEE Transactions on Power Systems}, vol.~32, no.~4, pp. 2604--2617, 2016.

\bibitem{npcc}
{Northeast Power Coordinating Council}, ``Npcc regional reliability reference directory \# 7 remedial action schemes.''

\bibitem{hedman2011review}
K.~W. Hedman, S.~S. Oren, and R.~P. O'Neill, ``A review of transmission switching and network topology optimization,'' in \emph{2011 IEEE power and energy society general meeting}.\hskip 1em plus 0.5em minus 0.4em\relax IEEE, 2011, pp. 1--7.

\bibitem{abdi2016corrective}
M.~Abdi-Khorsand, M.~Sahraei-Ardakani, and Y.~M. Al-Abdullah, ``Corrective transmission switching for n-1-1 contingency analysis,'' \emph{IEEE Transactions on Power Systems}, vol.~32, no.~2, pp. 1606--1615, 2016.

\bibitem{pjmsps}
{System Planning Division Transmission Planning Department PJM}, ``manual 07: pjm protection standards.''

\bibitem{salkuti2018congestion}
S.~R. Salkuti, ``Congestion management using optimal transmission switching,'' \emph{IEEE Systems Journal}, vol.~12, no.~4, pp. 3555--3564, 2018.

\bibitem{fisher2008optimal}
E.~B. Fisher, R.~P. O'Neill, and M.~C. Ferris, ``Optimal transmission switching,'' \emph{IEEE Transactions on Power Systems}, vol.~23, no.~3, pp. 1346--1355, 2008.

\bibitem{henneaux2015probabilistic}
P.~Henneaux and D.~S. Kirschen, ``Probabilistic security analysis of optimal transmission switching,'' \emph{IEEE Transactions on Power Systems}, vol.~31, no.~1, pp. 508--517, 2015.

\bibitem{khanabadi2012optimal}
M.~Khanabadi, H.~Ghasemi, and M.~Doostizadeh, ``Optimal transmission switching considering voltage security and n-1 contingency analysis,'' \emph{IEEE Transactions on Power Systems}, vol.~28, no.~1, pp. 542--550, 2012.

\bibitem{hedman2011optimal}
K.~W. Hedman, S.~S. Oren, and R.~P. O’Neill, ``Optimal transmission switching: economic efficiency and market implications,'' \emph{Journal of Regulatory Economics}, vol.~40, no.~2, pp. 111--140, 2011.

\bibitem{wang2016igbt}
H.~Wang and K.-W. Ma, ``Igbt technology for future high-power vsc-hvdc applications,'' in \emph{12th IET International Conference on AC and DC Power Transmission (ACDC 2016)}, 2016, pp. 1--6.

\bibitem{shenai2015invention}
K.~Shenai, ``The invention and demonstration of the igbt [a look back],'' \emph{ieee Power electronics magazine}, vol.~2, no.~2, pp. 12--16, 2015.

\bibitem{danielsson2015transmission}
J.~Danielsson, S.~Patel, J.~Pan, and R.~Nuqui, ``Transmission grid reinforcement with embedded vsc-hvdc,'' in \emph{Proc. CIGRE US National Committee 2015-Grid of the Future Symposium, Chicago, USA}, 2015, pp. 1--7.

\bibitem{bnef2017electronhighways}
{Bloomberg NEF}, ``{Technologies for High Voltage Transmission},'' {BNEF}, Tech. Rep., 2017.

\bibitem{usdoe2020advancedtransmission}
{U.S. Department of Energy}, ``{Advanced Transmission Technologies},'' {U.S. Department of Energy}, Tech. Rep., December 2020.

\bibitem{gu2018partial}
X.~Gu, S.~He, Y.~Xu, Y.~Yan, S.~Hou, and M.~Fu, ``Partial discharge detection on 320 kv vsc-hvdc xlpe cable with artificial defects under dc voltage,'' \emph{IEEE Transactions on Dielectrics and Electrical Insulation}, vol.~25, no.~3, pp. 939--946, 2018.

\bibitem{shu2018research}
Y.~Shu and W.~Chen, ``Research and application of uhv power transmission in china,'' \emph{High voltage}, vol.~3, no.~1, pp. 1--13, 2018.

\bibitem{chen2018variable}
G.~Chen, X.~Zhou, and R.~Chen, \emph{Variable frequency transformers for large scale power systems interconnection: theory and applications}.\hskip 1em plus 0.5em minus 0.4em\relax John Wiley \& Sons, 2018.

\bibitem{long2007hvdc}
W.~Long and S.~Nilsson, ``Hvdc transmission: yesterday and today,'' \emph{IEEE Power and Energy Magazine}, vol.~5, no.~2, pp. 22--31, 2007.

\bibitem{korompili2016review}
A.~Korompili, Q.~Wu, and H.~Zhao, ``Review of vsc hvdc connection for offshore wind power integration,'' \emph{Renewable and Sustainable Energy Reviews}, vol.~59, pp. 1405--1414, 2016.

\bibitem{sun2017renewable}
J.~Sun, M.~Li, Z.~Zhang, T.~Xu, J.~He, H.~Wang, and G.~Li, ``Renewable energy transmission by hvdc across the continent: system challenges and opportunities,'' \emph{CSEE Journal of Power and Energy Systems}, vol.~3, no.~4, pp. 353--364, 2017.

\bibitem{watson2020overview}
N.~R. Watson and J.~D. Watson, ``An overview of {HVDC} technology,'' \emph{Energies}, vol.~13, no.~17, p. 4342, 2020.

\bibitem{wen2017enhancing}
Y.~Wen, C.~Chung, and X.~Ye, ``Enhancing frequency stability of asynchronous grids interconnected with hvdc links,'' \emph{IEEE Transactions on Power Systems}, vol.~33, no.~2, pp. 1800--1810, 2017.

\bibitem{wendt20152030}
V.~E. Wendt, ``2030 15\% interconnection target: Challenges \& solutions for a timely project implementation,'' \emph{Vienna: EuropaCable}, 2015.

\bibitem{rao2020frequency}
H.~Rao, W.~Wu, T.~Mao, B.~Zhou, C.~Hong, Y.~Liu, and X.~Wu, ``Frequency control at the power sending side for hvdc asynchronous interconnections between yunnan power grid and the rest of csg,'' \emph{CSEE Journal of Power and Energy Systems}, vol.~7, no.~1, pp. 105--113, 2020.

\bibitem{alassi2019hvdc}
A.~Alassi, S.~Ba{\~n}ales, O.~Ellabban, G.~Adam, and C.~MacIver, ``Hvdc transmission: technology review, market trends and future outlook,'' \emph{Renewable and Sustainable Energy Reviews}, vol. 112, pp. 530--554, 2019.

\bibitem{li2016research}
B.~Li, T.~Liu, W.~Xu, Q.~Li, Y.~Zhang, Y.~Li, and X.~Y. Li, ``Research on technical requirements of line-commutated converter-based high-voltage direct current participating in receiving end ac system's black start,'' \emph{IET Generation, Transmission \& Distribution}, vol.~10, no.~9, pp. 2071--2078, 2016.

\bibitem{xiong2020modeling}
L.~Xiong, X.~Liu, Y.~Liu, and F.~Zhuo, ``Modeling and stability issues of voltage-source converter dominated power systems: A review,'' \emph{CSEE Journal of Power and Energy Systems}, 2020.

\bibitem{rodriguez2017multi}
P.~Rodriguez and K.~Rouzbehi, ``Multi-terminal dc grids: challenges and prospects,'' \emph{Journal of Modern Power Systems and Clean Energy}, vol.~5, no.~4, pp. 515--523, 2017.

\bibitem{xiang2016cost}
X.~Xiang, M.~M.~C. Merlin, and T.~C. Green, ``Cost analysis and comparison of hvac, lfac and hvdc for offshore wind power connection,'' in \emph{12th IET International Conference on AC and DC Power Transmission (ACDC 2016)}, 2016, pp. 1--6.

\bibitem{chen2017analysis}
Z.~Chen, Z.~Yu, X.~Zhang, T.~Wei, G.~Lyu, L.~Qu, Y.~Huang, and R.~Zeng, ``Analysis and experiments for igbt, iegt, and igct in hybrid dc circuit breaker,'' \emph{IEEE Transactions on Industrial Electronics}, vol.~65, no.~4, pp. 2883--2892, 2017.

\bibitem{montanari2018criteria}
G.~C. Montanari, P.~H. Morshuis, M.~Zhou, G.~C. Stevens, A.~S. Vaughan, Z.~Han, and D.~Li, ``Criteria influencing the selection and design of hv and uhv dc cables in new network applications,'' \emph{High Voltage}, vol.~3, no.~2, pp. 90--95, 2018.

\bibitem{rahman2021comparison}
S.~Rahman, I.~Khan, H.~I. Alkhammash, and M.~F. Nadeem, ``A comparison review on transmission mode for onshore integration of offshore wind farms: Hvdc or hvac,'' \emph{Electronics}, vol.~10, no.~12, p. 1489, 2021.

\bibitem{mihalivc2004transient}
R.~Mihali{\v{c}} and U.~Gabrijel, ``Transient stability assessment of systems comprising phase-shifting facts devices by direct methods,'' \emph{International journal of electrical power \& energy systems}, vol.~26, no.~6, pp. 445--453, 2004.

\bibitem{sajjadi2021svpwm}
M.~Sajjadi and Y.~Majd, ``Svpwm-based dual active filter for distribution system power quality improvement,'' in \emph{2021 6th International Conference on Power and Renewable Energy (ICPRE)}.\hskip 1em plus 0.5em minus 0.4em\relax IEEE, 2021, pp. 592--597.

\bibitem{siemens_pst}
\BIBentryALTinterwordspacing
{Siemens Energy}, ``{Phase-shifting transformers},'' 2022, accessed = 2022-12-02. [Online]. Available: \url{https://www.siemens-energy.com/global/en/offerings/power-transmission/portfolio/transformers/phase-shifting-transformers.html}
\BIBentrySTDinterwordspacing

\bibitem{huang2003feature}
C.-N. Huang, ``Feature analysis of power flows based on the allocations of phase-shifting transformers,'' \emph{IEEE transactions on power systems}, vol.~18, no.~1, pp. 266--272, 2003.

\bibitem{rasolomampionona2011interaction}
D.~Rasolomampionona and S.~Anwar, ``Interaction between phase shifting transformers installed in the tie-lines of interconnected power systems and automatic frequency controllers,'' \emph{International Journal of Electrical Power \& Energy Systems}, vol.~33, no.~8, pp. 1351--1360, 2011.

\bibitem{verboomen2008use}
J.~Verboomen, G.~Papaefthymiou, W.~Kling, and L.~Van~der Sluis, ``Use of phase shifting transformers for minimising congestion risk,'' in \emph{Proceedings of the 10th International Conference on Probablistic Methods Applied to Power Systems}.\hskip 1em plus 0.5em minus 0.4em\relax IEEE, 2008, pp. 1--6.

\bibitem{ippolito2004selection}
L.~Ippolito and P.~Siano, ``Selection of optimal number and location of thyristor-controlled phase shifters using genetic based algorithms,'' \emph{IEE Proceedings-Generation, Transmission and Distribution}, vol. 151, no.~5, pp. 630--637, 2004.

\bibitem{urresti2020pre}
E.~Urresti-Padr{\'o}n, M.~Jakubek, W.~Jaworski, and M.~K{\l}os, ``Pre-selection of the optimal sitting of phase-shifting transformers based on an optimization problem solved within a coordinated cross-border congestion management process,'' \emph{Energies}, vol.~13, no.~14, p. 3748, 2020.

\bibitem{sidea2017sizing}
D.~O. Sidea, L.~Toma, and M.~Eremia, ``Sizing a phase shifting transformer for congestion management in high wind generation areas,'' in \emph{2017 IEEE Manchester PowerTech}.\hskip 1em plus 0.5em minus 0.4em\relax IEEE, 2017, pp. 1--6.

\bibitem{korab2016impact}
R.~Korab and R.~Owczarek, ``Impact of phase shifting transformers on cross-border power flows in the central and eastern europe region,'' \emph{Bulletin of the Polish Academy of Sciences. Technical Sciences}, vol.~64, no.~1, 2016.

\bibitem{nikoobakht2019stochastic}
A.~Nikoobakht, J.~Aghaei, R.~Khatami, E.~Mahboubi-Moghaddam, and M.~Parvania, ``Stochastic flexible transmission operation for coordinated integration of plug-in electric vehicles and renewable energy sources,'' \emph{Applied energy}, vol. 238, pp. 225--238, 2019.

\bibitem{fernandez2016review}
E.~Fernandez, I.~Albizu, M.~Bedialauneta, A.~Mazon, and P.~T. Leite, ``Review of dynamic line rating systems for wind power integration,'' \emph{Renewable and Sustainable Energy Reviews}, vol.~53, pp. 80--92, 2016.

\bibitem{wallnerstrom2014impact}
C.~J. Wallnerstr{\"o}m, Y.~Huang, and L.~S{\"o}der, ``Impact from dynamic line rating on wind power integration,'' \emph{IEEE Transactions on Smart Grid}, vol.~6, no.~1, pp. 343--350, 2014.

\bibitem{michiorri2015forecasting}
A.~Michiorri, H.-M. Nguyen, S.~Alessandrini, J.~B. Bremnes, S.~Dierer, E.~Ferrero, B.-E. Nygaard, P.~Pinson, N.~Thomaidis, and S.~Uski, ``Forecasting for dynamic line rating,'' \emph{Renewable and sustainable energy reviews}, vol.~52, pp. 1713--1730, 2015.

\bibitem{ferc2019dlr}
\BIBentryALTinterwordspacing
{Federal Energy Regulatory Commission (FERC)}, ``Managing transmission line ratings,'' 2019. [Online]. Available: \url{https://www.ferc.gov/sites/default/files/2020-05/tran-line-ratings.pdf}
\BIBentrySTDinterwordspacing

\bibitem{inl-dynamic-line-rating}
{Idaho National Laboratory}, ``{Dynamic Line Rating Overview},'' 2021.

\bibitem{hur2009high}
K.~Hur, M.~Boddeti, N.~Sarma, J.~Dumas, J.~Adams, and S.-K. Chai, ``High-wire act,'' \emph{IEEE Power and Energy Magazine}, vol.~8, no.~1, pp. 37--45, 2009.

\bibitem{li2019day}
Y.~Li, B.~Hu, K.~Xie, L.~Wang, Y.~Xiang, R.~Xiao, and D.~Kong, ``Day-ahead scheduling of power system incorporating network topology optimization and dynamic thermal rating,'' \emph{IEEE Access}, vol.~7, pp. 35\,287--35\,301, 2019.

\bibitem{energy-dynamic-line-rating}
{U.S. Department of Energy}, ``{Improving Efficiency with Dynamic Line Ratings},'' 2017.

\bibitem{potomac-economics-miso-report}
{Potomac Economics}, ``{2020 MISO State of the Market Report},'' \url{https://www.potomaceconomics.com/wp-content/uploads/2021/05/2020-MISO-SOM_Report_Body_Compiled_Final_rev-6-1-21.pdf}, 2021.

\bibitem{caiso-managing-line-ratings}
{California Independent System Operator}, ``{Comments on Managing Transmission Line Ratings},'' \url{https://www.caiso.com/Documents/Mar22-2021-Comments-ManagingTransmissionLineRatings-RM20-16.pdf}, 2021.

\bibitem{iso-ne-order-881}
{ISO New England}, ``{ISO-NE Order 881 Presentation},'' \url{https://www.iso-ne.com/static-assets/documents/2022/04/a4_order_881_presentation.pdf}, 2022.

\bibitem{nyiso-ferc-order-881}
{New York Independent System Operator (NYISO)}, ``{NYISO - FERC Order 881},'' \url{https://www.nyiso.com/documents/20142/29177064/03162022%20NYISO%20-%20FERC%20Order%20881%20v2.pdf/38819aac-890f-5412-a945-ad6817c6676e}, 2022.

\bibitem{spp-order-881-compliance}
{Southwest Power Pool (SPP)}, ``{Order No. 881 Compliance Filing to Implement Transmission Line Ratings},'' \url{https://www.spp.org/documents/67491/20220712_order%20no.%20881%20compliance%20filing%20to%20implement%20transmission%20line%20ratings_er22-2339-000.pdf}, 2022.

\bibitem{ferc-order-1000}
{Federal Energy Regulatory Commission (FERC)}, ``{Order No. 1000: Transmission Planning and Cost Allocation by Transmission Owning and Operating Public Utilities},'' Federal Register, vol. 76, no. 125, pp. 38402-38475, 14 July 2011, 2011.

\bibitem{shao2019low}
C.~Shao, Y.~Ding, and J.~Wang, ``A low-carbon economic dispatch model incorporated with consumption-side emission penalty scheme,'' \emph{Applied Energy}, vol. 238, pp. 1084--1092, 2019.

\bibitem{sun2017analysis}
Y.~Sun, C.~Kang, Q.~Xia, Q.~Chen, N.~Zhang, and Y.~Cheng, ``Analysis of transmission expansion planning considering consumption-based carbon emission accounting,'' \emph{Applied energy}, vol. 193, pp. 232--242, 2017.

\bibitem{seddighi2015integrated}
A.~H. Seddighi and A.~Ahmadi-Javid, ``Integrated multiperiod power generation and transmission expansion planning with sustainability aspects in a stochastic environment,'' \emph{Energy}, vol.~86, pp. 9--18, 2015.

\bibitem{lokeshgupta2018multi}
B.~Lokeshgupta and S.~Sivasubramani, ``Multi-objective dynamic economic and emission dispatch with demand side management,'' \emph{International Journal of Electrical Power \& Energy Systems}, vol.~97, pp. 334--343, 2018.

\bibitem{park2018stochastic}
H.~Park, Y.~G. Jin, and J.-K. Park, ``Stochastic security-constrained unit commitment with wind power generation based on dynamic line rating,'' \emph{International Journal of Electrical Power \& Energy Systems}, vol. 102, pp. 211--222, 2018.

\bibitem{peker2018benefits}
M.~Peker, A.~S. Kocaman, and B.~Y. Kara, ``Benefits of transmission switching and energy storage in power systems with high renewable energy penetration,'' \emph{Applied Energy}, vol. 228, pp. 1182--1197, 2018.

\bibitem{jabarnejad2020facilitating}
M.~Jabarnejad, ``Facilitating emission reduction using the dynamic line switching and rating,'' \emph{Electric Power Systems Research}, vol. 189, p. 106600, 2020.

\bibitem{mirzapour2023transmission}
O.~Mirzapour, X.~Rui, and M.~Sahraei-Ardakani, ``Transmission impedance control impacts on carbon emissions and renewable energy curtailment,'' \emph{Energy}, vol. 278, p. 127741, 2023.

\bibitem{ela2014evolution}
E.~Ela, M.~Milligan, A.~Bloom, A.~Botterud, A.~Townsend, and T.~Levin, ``Evolution of wholesale electricity market design with increasing levels of renewable generation,'' \emph{NREL}, 2014.

\bibitem{wu2019pool}
Y.~Wu, M.~Barati, and G.~J. Lim, ``A pool strategy of microgrid in power distribution electricity market,'' \emph{IEEE Transactions on Power Systems}, vol.~35, no.~1, pp. 3--12, 2019.

\bibitem{sahraei2018merchant}
M.~Sahraei-Ardakani, ``Merchant power flow controllers,'' \emph{Energy Economics}, vol.~74, pp. 878--885, 2018.

\bibitem{sahraei2015transfer}
M.~Sahraei-Ardakani and S.~A. Blumsack, ``Transfer capability improvement through market-based operation of series facts devices,'' \emph{IEEE Transactions on Power Systems}, vol.~31, no.~5, pp. 3702--3714, 2015.

\bibitem{sahraei2012active}
M.~Sahraei-Ardakani and S.~Blumsack, ``Active participation of facts devices in wholesale electricity markets,'' in \emph{Proc. of 31 USAEE North American Conference}.\hskip 1em plus 0.5em minus 0.4em\relax Citeseer, 2012.

\bibitem{sahraei2012market}
------, ``Market equilibrium for dispatchable transmission using fact devices,'' in \emph{2012 IEEE Power and Energy Society General Meeting}.\hskip 1em plus 0.5em minus 0.4em\relax IEEE, 2012, pp. 1--6.

\bibitem{sahraei2013marginal}
M.~Sahraei-Ardakani and S.~A. Blumsack, ``Marginal value of facts devices in transmission-constrained electricity markets,'' in \emph{2013 IEEE Power \& Energy Society General Meeting}.\hskip 1em plus 0.5em minus 0.4em\relax IEEE, 2013, pp. 1--5.

\bibitem{fuller2012fast}
J.~D. Fuller, R.~Ramasra, and A.~Cha, ``Fast heuristics for transmission-line switching,'' \emph{IEEE Transactions on Power Systems}, vol.~27, no.~3, pp. 1377--1386, 2012.

\bibitem{pjm_future_report}
\BIBentryALTinterwordspacing
{PJM Planning Division}, ``{Grid of the Future: PJM’s Regional Planning Perspective},'' May 2022, last accessed 5 February 2023. [Online]. Available: \url{https://pjm.com/-/media/library/reports-notices/special-reports/2022/20220510-grid-of-the-future-pjms-regional-planning-perspective.ashx}
\BIBentrySTDinterwordspacing

\bibitem{ferc881web}
\BIBentryALTinterwordspacing
M.~O’Driscoll, ``{FERC Rule} to improve transmission line ratings will help lower transmission costs,'' December 2021, last accessed 18 January 2023. [Online]. Available: \url{https://www.ferc.gov/news-events/news/ferc-rule-improve-transmission-line-ratings-will-help-lower-transmission-costs}
\BIBentrySTDinterwordspacing

\bibitem{fercworkshop}
\BIBentryALTinterwordspacing
Workshop to discuss certain performance-based ratemaking approaches. Accessed: 2022-11-19. [Online]. Available: \url{https://www.ferc.gov/news-events/events/workshop-discuss-certain-performance-based-ratemaking-approaches-09102021}
\BIBentrySTDinterwordspacing

\bibitem{pjm_problem_statement}
\BIBentryALTinterwordspacing
{PJM}, ``Potential additional issues relating to the implementation of dynamic line ratings,'' August 2022, last accessed 5 July 2023. [Online]. Available: \url{https://www.pjm.com/-/media/committees-groups/task-forces/dlrtf/postings/dlrtf-problem-statement.ashx}
\BIBentrySTDinterwordspacing

\bibitem{isone_881_compliance}
\BIBentryALTinterwordspacing
G.~Jesmer, ``{FERC Order No. 881 Compliance},'' March 2022, last accessed 5 July 2023. [Online]. Available: \url{https://www.iso-ne.com/static-assets/documents/2022/03/a4_order_no_881_compliance_incorporation_of_ambient_adjusted_line_ratings.pdf}
\BIBentrySTDinterwordspacing

\bibitem{rogers2008some}
K.~Rogers and T.~J. Overbye, ``Some applications of distributed flexible ac transmission system (d-facts) devices in power systems,'' in \emph{2008 40th North American Power Symposium}.\hskip 1em plus 0.5em minus 0.4em\relax IEEE, 2008, pp. 1--8.

\bibitem{soroudi2021controllable}
A.~Soroudi, ``Controllable transmission networks under demand uncertainty with modular facts,'' \emph{International Journal of Electrical Power \& Energy Systems}, vol. 130, p. 106978, 2021.

\bibitem{pjm_benefits}
\BIBentryALTinterwordspacing
{PJM}, ``The benefits of the {PJM} transmission system,'' April 2019, last accessed 19 January 2023. [Online]. Available: \url{https://www.pjm.com/-/media/library/reports-notices/special-reports/2019/the-benefits-of-the-pjm-transmission-system.pdf}
\BIBentrySTDinterwordspacing

\bibitem{isone_regional_2021}
{ISO New England Inc.}, ``{2021 Regional System Plan}.''

\bibitem{iso-ne_tranplan}
\BIBentryALTinterwordspacing
------, ``Transmission planning technical guide,'' February 2022. [Online]. Available: \url{https://www.iso-ne.com/static-assets/documents/2022/02/transmission_planning_technical_guide_rev7_2.pdf}
\BIBentrySTDinterwordspacing

\bibitem{caiso_tranplan}
\BIBentryALTinterwordspacing
{California ISO}, ``2021-2022 transmission plan,'' March 2022, last accessed 30 January 2023. [Online]. Available: \url{http://www.caiso.com/InitiativeDocuments/ISOBoardApproved-2021-2022TransmissionPlan.pdf}
\BIBentrySTDinterwordspacing

\bibitem{pjm_teac}
\BIBentryALTinterwordspacing
{PJM Transmission Expansion Advisory Committee}, ``Reliability analysis update,'' January 2018, last accessed 30 January 2023. [Online]. Available: \url{https://www.pjm.com/-/media/committees-groups/committees/teac/20180111/20180111-reliability-analysis-update.ashx}
\BIBentrySTDinterwordspacing

\bibitem{nypa_mssc_press}
\BIBentryALTinterwordspacing
M.~Balaban, ``{NYPA} completes electric grid project to improve reliability and bring more renewable energy downstate,'' June 2016, last accessed 30 January 2023. [Online]. Available: \url{https://www.nypa.gov/news/press-releases/2016/20160614-marcy-south-completed}
\BIBentrySTDinterwordspacing

\bibitem{doe_advanced_transmission}
\BIBentryALTinterwordspacing
{U.S. Department of Energy}, ``Advanced transmission technologies,'' December 2020, last accessed 30 January 2023. [Online]. Available: \url{https://www.energy.gov/sites/prod/files/2021/02/f82/Advanced%20Transmission%20Technologies%20Report%20-%20final%20as%20of%2012.3%20-%20FOR%20PUBLIC.pdf}
\BIBentrySTDinterwordspacing

\bibitem{ge_seriescompensation}
\BIBentryALTinterwordspacing
{{GE Grid Solutions}}. Series compensation systems. Last accessed 6 February 2023. [Online]. Available: \url{https://www.gegridsolutions.com/products/brochures/powerD_vtf/SeriesCompensation_GEA12785C_LR.pdf}
\BIBentrySTDinterwordspacing

\bibitem{hitachi_tcsc}
\BIBentryALTinterwordspacing
{{Hitachi Energy}}. Thyristor controlled series compensation. Last accessed 6 February 2023. [Online]. Available: \url{https://www.hitachienergy.com/us/en/products-and-solutions/facts/thyristor-controlled-series-compensation}
\BIBentrySTDinterwordspacing

\bibitem{siemens_facts}
\BIBentryALTinterwordspacing
{{Siemens Energy}}. Flexible {AC} transmission systems. Last accessed 6 February 2023. [Online]. Available: \url{https://www.siemens-energy.com/global/en/offerings/power-transmission/portfolio/flexible-ac-transmission-systems.html}
\BIBentrySTDinterwordspacing

\bibitem{siemens_upfcplus}
{Siemens Energy}, ``{UPFC PLUS}.''

\bibitem{smartwires_smartvalve}
\BIBentryALTinterwordspacing
{{SmartValve$^{TM}$}}. {{SmartWires}}. Last accessed 6 February 2023. [Online]. Available: \url{https://www.smartwires.com/smartvalve/}
\BIBentrySTDinterwordspacing

\bibitem{Brattle_advancedtechnologies}
\BIBentryALTinterwordspacing
T.~B. Tsuchida and R.~Gramlich, ``Improving transmission operation with advanced technologies: A review of deployment experience and analysis of incentives,'' June 2019, last accessed 6 February 2023. [Online]. Available: \url{https://www.brattle.com/wp-content/uploads/2021/05/16634_improving_transmission_operating_with_advanced_technologies.pdf}
\BIBentrySTDinterwordspacing

\bibitem{ruiz_ferc_switching}
\BIBentryALTinterwordspacing
P.~A. Ruiz, M.~Caramanis, E.~Goldis, X.~Li, K.~Patel, R.~Philbrick, A.~Rudkevich, R.~Tabors, and B.~Tsuchida, ``Transmission topology optimization,'' June 2016, last accessed 6 February 2023. [Online]. Available: \url{https://cms.ferc.gov/sites/default/files/2020-05/20160629114654-2%2520-%2520PRuiz%2520FERCTechConf%252028Jun2016_FINAL_2.pdf}
\BIBentrySTDinterwordspacing

\bibitem{ruiz_ferc_switching_spp_ercot}
\BIBentryALTinterwordspacing
P.~A. Ruiz, J.~Caspary, and L.~Butler, ``Transmission topology optimization case studies in {SPP} and {ERCOT},'' June 2020, last accessed 6 February 2023. [Online]. Available: \url{https://www.ferc.gov/sites/default/files/2020-06/W3-1_Ruiz_et_al.pdf}
\BIBentrySTDinterwordspacing

\bibitem{PJM_switching_solutions}
\BIBentryALTinterwordspacing
{{PJM}}. {Swithcing solutions}. Last accessed 6 February 2023. [Online]. Available: \url{https://www.pjm.com/markets-and-operations/etools/oasis/system-information/switching-solutions}
\BIBentrySTDinterwordspacing

\bibitem{newgrid_arpae}
\BIBentryALTinterwordspacing
{{Advanced Research Projects Agency - Energy}}. {{NewGrid}}. Last accessed 6 February 2023. [Online]. Available: \url{https://arpa-e.energy.gov/technologies/scaleup-launch-pad-2020/newgrid}
\BIBentrySTDinterwordspacing

\bibitem{ruiz_switching_spp_pilot}
\BIBentryALTinterwordspacing
P.~A. Ruiz and J.~Caspary, ``{SPP} transmission topology optimization pilot,'' March 2019, last accessed 5 July 2023. [Online]. Available: \url{https://watt-transmission.org/wp-content/uploads/2019/03/spp-transmission-topology-optimization-pilot-efficient-congestion-management-and-overload-mitigation-through-system-reconfigurations-.pdf}
\BIBentrySTDinterwordspacing

\bibitem{doe_improving_dlr}
\BIBentryALTinterwordspacing
{U.S. Department of Energy}, ``Improving efficiency with dynamic line ratings,'' last accessed 31 January 2023. [Online]. Available: \url{https://www.energy.gov/sites/prod/files/2017/01/f34/NYPA_Improving-Efficiency-Dynamic-Line-Ratings.pdf}
\BIBentrySTDinterwordspacing

\bibitem{pjm_dlr_webnews}
\BIBentryALTinterwordspacing
{PJM Inside Lines}, ``{PJM} facilitating dynamic line rating implementation,'' March 2022, last accessed 18 January 2023. [Online]. Available: \url{https://insidelines.pjm.com/pjm-facilitating-dynamic-line-rating-implementation/}
\BIBentrySTDinterwordspacing

\bibitem{ferc_dlr}
\BIBentryALTinterwordspacing
{Federal Energy Regulatory Commission}, ``Managing transmission line ratings,'' August 2019, last accessed 23 January 2023. [Online]. Available: \url{https://www.ferc.gov/sites/default/files/2020-05/tran-line-ratings.pdf}
\BIBentrySTDinterwordspacing

\bibitem{doe_dlr}
\BIBentryALTinterwordspacing
{U.S. Department of Energy}, ``Dynamic line rating,'' June 2019, last accessed 19 January 2023. [Online]. Available: \url{https://www.energy.gov/sites/default/files/2021/03/f83/DLR%20Report%20-%20June%202019%20final%20-%20FOR%20PUBLIC%20USE.pdf}
\BIBentrySTDinterwordspacing

\bibitem{pjm_aar_setting}
\BIBentryALTinterwordspacing
D.~Hislop, ``Managing transmission line ratings,'' February 2022, last accessed 31 January 2023. [Online]. Available: \url{https://www.pjm.com/-/media/committees-groups/committees/oc/2022/20220210/20220210-item-09-managing-transmission-line-ratings-order-no-881-compliance-filing-rm20-16-000-presentation.ashx}
\BIBentrySTDinterwordspacing

\bibitem{linevision_dlr}
\BIBentryALTinterwordspacing
{LineVision}. Technology. Last accessed 31 January 2023. [Online]. Available: \url{https://www.linevisioninc.com/technology}
\BIBentrySTDinterwordspacing

\bibitem{Sumo_operato}
\BIBentryALTinterwordspacing
Operato. {SUMO}. Last accessed 31 January 2023. [Online]. Available: \url{https://www.operato.eu/sumo}
\BIBentrySTDinterwordspacing

\bibitem{Sumo_smartwires}
\BIBentryALTinterwordspacing
{{Smart Wires}}. {Smart Wires extends commercial offering by partnering with software provider Operato}. Last accessed 31 January 2023. [Online]. Available: \url{https://www.smartwires.com/2022/08/10/smart-wires-partners-with-software-provider/}
\BIBentrySTDinterwordspacing

\bibitem{cook2018phase}
B.~Cook, M.~J. Thompson, K.~Garg, and M.~Malichkar, ``Phase-shifting transformer control and protection settings verification,'' in \emph{2018 71st Annual Conference for Protective Relay Engineers (CPRE)}.\hskip 1em plus 0.5em minus 0.4em\relax IEEE, 2018, pp. 1--15.

\bibitem{caiso_business}
\BIBentryALTinterwordspacing
{California ISO}, ``Business requirements specification,'' September 2017. [Online]. Available: \url{https://www.caiso.com/Documents/BusinessRequirementsSpecification-PhaseShifterModeling.pdf}
\BIBentrySTDinterwordspacing

\bibitem{nyiso_par}
\BIBentryALTinterwordspacing
{New York ISO}, ``Manual 11 day-ahead scheduling manual,'' December 2022, last accessed 29 January 2023. [Online]. Available: \url{https://www.nyiso.com/documents/20142/2923301/dayahd_schd_mnl.pdf/0024bc71-4dd9-fa80-a816-f9f3e26ea53a}
\BIBentrySTDinterwordspacing

\bibitem{hui2012wind}
H.~Hui, C.-N. Yu, R.~Surendran, F.~Gao, and S.~Moorty, ``Wind generation scheduling and coordination in ercot nodal market,'' in \emph{2012 IEEE Power and Energy Society General Meeting}.\hskip 1em plus 0.5em minus 0.4em\relax IEEE, 2012, pp. 1--8.

\bibitem{caiso_tranplan_process}
\BIBentryALTinterwordspacing
{California ISO}, ``{CAISO} transmission planning process,'' March 2022, last accessed 1 February 2023. [Online]. Available: \url{http://www.caiso.com/Documents/Presentation-CaliforniaISOAnnualInterregionalInformationMar042022.pdf}
\BIBentrySTDinterwordspacing

\bibitem{nyiso_internal_controllable}
\BIBentryALTinterwordspacing
A.~Myott, ``Internal controllable lines,'' February 2022, last accessed 1 February 2023. [Online]. Available: \url{https://www.nyiso.com/documents/20142/28227906/Internal%20Controllable%20Lines_02032022_FINAL.pdf/6ea8f352-aa78-2888-5f1b-9a5e0075da58}
\BIBentrySTDinterwordspacing

\bibitem{CPNY_HVDC}
\BIBentryALTinterwordspacing
{{Clean Path NY}}. {{Environmental Benefits}}. Last accessed 2 February 2023. [Online]. Available: \url{https://www.cleanpathny.com/environmental-benefits}
\BIBentrySTDinterwordspacing

\bibitem{CHPE_HVDC}
\BIBentryALTinterwordspacing
{{Champlain Hudson Power Express}}. {{The Technology}}. Last accessed 2 February 2023. [Online]. Available: \url{https://chpexpress.com/project-overview/the-technology/}
\BIBentrySTDinterwordspacing

\bibitem{pjm_soogreen}
\BIBentryALTinterwordspacing
S.~Frenkel, ``{SOO Green HVDC Link},'' May 2020, last accessed 2 February 2023. [Online]. Available: \url{https://www.pjm.com/-/media/committees-groups/committees/mrc/2020/20200522-hvdc/20200522-item-03-soo-green-hvdc-link-presentation.ashx}
\BIBentrySTDinterwordspacing

\bibitem{pjm_hvdcstf_report}
\BIBentryALTinterwordspacing
{PJM}, ``{High Voltage Direct Current Senior Task Force (HVDCSTF) Final Report},'' December 2021, last accessed 2 February 2023. [Online]. Available: \url{https://www.pjm.com/-/media/committees-groups/task-forces/hvdcstf/postings/hvdc-final-report.ashx}
\BIBentrySTDinterwordspacing

\bibitem{ge_hvdc_brochure}
\BIBentryALTinterwordspacing
{GE Grid Solutions}, ``High voltage direct current systems,'' 2016, last accessed 2 February 2023. [Online]. Available: \url{https://resources.gegridsolutions.com/hvdc/hvdc-systems-brochure}
\BIBentrySTDinterwordspacing

\bibitem{HVDC_hitachi}
\BIBentryALTinterwordspacing
{{Hitachi Energy}}. {{HVDC}}. Last accessed 2 February 2023. [Online]. Available: \url{https://www.hitachienergy.com/us/en/products-and-solutions/hvdc}
\BIBentrySTDinterwordspacing

\bibitem{alrasheedi2023unit}
A.~F. Alrasheedi, K.~A. Alnowibet, and A.~M. Alshamrani, ``A unit commitment based-co-optimization of generation and transmission expansion planning to mitigate market power,'' \emph{Electric Power Systems Research}, vol. 214, p. 108860, 2023.

\bibitem{opgrand2019role}
I.~Opgrand and J.~Jeffrey, ``The role of auction revenue rights in markets for financial transmission rights,'' \emph{Bid}, vol.~1, p.~1, 2019.

\bibitem{kuosmanen2020capital}
T.~Kuosmanen and T.~Nguyen, ``Capital bias in the nordic revenue cap regulation: Averch-johnson critique revisited,'' \emph{Energy Policy}, vol. 139, p. 111355, 2020.

\bibitem{aazami2012comprehensive}
R.~Aazami, M.~R. Haghifam, and M.~Doostizadeh, ``Comprehensive modeling of flexible transmission services in stochastic joint energy and spinning reserve market,'' \emph{International Journal of Electrical Power \& Energy Systems}, vol.~43, no.~1, pp. 1354--1362, 2012.

\bibitem{staudt2021merchant}
P.~Staudt and S.~S. Oren, ``Merchant transmission in single-price electricity markets with cost-based redispatch,'' \emph{Energy Economics}, vol. 104, p. 105610, 2021.

\bibitem{joskow2020competition}
P.~L. Joskow, ``Competition for electric transmission projects in the usa: Ferc order 1000,'' \emph{Transmission Network Investment in Liberalized Power Markets}, pp. 275--322, 2020.

\bibitem{schulte2020vision}
R.~H. Schulte and F.~C. Fletcher, ``Why the vision of interregional electric transmission development in ferc order 1000 is not happening,'' \emph{The Electricity Journal}, vol.~33, no.~6, p. 106773, 2020.

\bibitem{matschoss2019german}
P.~Matschoss, B.~Bayer, H.~Thomas, and A.~Marian, ``The german incentive regulation and its practical impact on the grid integration of renewable energy systems,'' \emph{Renewable Energy}, vol. 134, pp. 727--738, 2019.

\bibitem{marques2022grid}
L.~Marques, A.~Sanjab, Y.~Mou, H.~Le~Cadre, and K.~Kessels, ``Grid impact aware tso-dso market models for flexibility procurement: Coordination, pricing efficiency, and information sharing,'' \emph{IEEE Transactions on Power Systems}, 2022.

\bibitem{haaberg2019fundamentals}
M.~H{\aa}berg, ``Fundamentals and recent developments in stochastic unit commitment,'' \emph{International Journal of Electrical Power \& Energy Systems}, vol. 109, pp. 38--48, 2019.

\bibitem{lumbreras2016new}
S.~Lumbreras and A.~Ramos, ``The new challenges to transmission expansion planning. survey of recent practice and literature review,'' \emph{Electric Power Systems Research}, vol. 134, pp. 19--29, 2016.

\bibitem{sahraei2016computationally}
M.~Sahraei-Ardakani and K.~W. Hedman, ``Computationally efficient adjustment of facts set points in dc optimal power flow with shift factor structure,'' \emph{IEEE Transactions on Power Systems}, vol.~32, no.~3, pp. 1733--1740, 2016.

\bibitem{kocuk2017new}
B.~Kocuk, S.~S. Dey, and X.~A. Sun, ``New formulation and strong misocp relaxations for ac optimal transmission switching problem,'' \emph{IEEE Transactions on Power Systems}, vol.~32, no.~6, pp. 4161--4170, 2017.

\bibitem{bai2016two}
Y.~Bai, H.~Zhong, Q.~Xia, and C.~Kang, ``A two-level approach to ac optimal transmission switching with an accelerating technique,'' \emph{IEEE Transactions on Power Systems}, vol.~32, no.~2, pp. 1616--1625, 2016.

\bibitem{vita2021evaluating}
V.~Vita, C.~Christodoulou, I.~Zafeiropoulos, I.~Gonos, M.~Asprou, and E.~Kyriakides, ``Evaluating the flexibility benefits of smart grid innovations in transmission networks,'' \emph{Applied Sciences}, vol.~11, no.~22, p. 10692, 2021.

\bibitem{eid2015aggregation}
C.~Eid, P.~Codani, Y.~Chen, Y.~Perez, and R.~Hakvoort, ``Aggregation of demand side flexibility in a smart grid: A review for european market design,'' in \emph{2015 12th International Conference on the European Energy Market (EEM)}.\hskip 1em plus 0.5em minus 0.4em\relax IEEE, 2015, pp. 1--5.

\bibitem{schachter2016critical}
J.~A. Schachter and P.~Mancarella, ``A critical review of real options thinking for valuing investment flexibility in smart grids and low carbon energy systems,'' \emph{Renewable and Sustainable Energy Reviews}, vol.~56, pp. 261--271, 2016.

\end{thebibliography}

\vfill

\end{document}